\documentclass[a4paper,11pt]{article}

\usepackage[margin = 1in]{geometry}
\usepackage[normalem]{ulem}
\usepackage{microtype,xcolor}
\usepackage[hidelinks]{hyperref}
\hypersetup{
    colorlinks,
    linkcolor={red!50!black},
    citecolor={blue!50!black},
    urlcolor={blue!80!black}
}
\usepackage{subcaption}
\usepackage{caption}
\usepackage{graphicx}
\usepackage{algorithm}
\usepackage{algpseudocode}
\usepackage[toc,page]{appendix}

\usepackage{amsmath,amsfonts,amssymb,bbm}
\usepackage{amsthm}
\usepackage{color}
\usepackage{complexity}
\usepackage{bm,enumitem} 
\newtheorem{definition}{Definition}[section]
\newtheorem{remark}{Remark}[section]
\newtheorem{claim}{Claim}[section]

\usepackage{aliascnt} %IMPORTANT For correct autoref with shared counters. Not necessary if counters are not shared.
\newtheorem{theorem}{Theorem}[section]
\newtheorem*{theorem*}{Theorem}

\newaliascnt{lemma}{theorem}
\newtheorem{lemma}[lemma]{Lemma}
\aliascntresetthe{lemma}

\newtheorem*{lemma*}{Lemma}

\newaliascnt{fact}{theorem}

\aliascntresetthe{fact}

\newtheorem*{fact*}{Fact}

\newaliascnt{observation}{theorem}

\aliascntresetthe{observation}

\newtheorem*{observation*}{Observation}

\newaliascnt{conjecture}{theorem}

\aliascntresetthe{conjecture}

\newtheorem*{conjecture*}{Conjecture}

\newaliascnt{corollary}{theorem}
\newtheorem{corollary}[corollary]{Corollary}
\aliascntresetthe{corollary}

\newtheorem*{corollary*}{Corollary}

\newaliascnt{proposition}{theorem}
\newtheorem{proposition}[proposition]{Proposition}
\aliascntresetthe{proposition}

\newtheorem*{proposition*}{Proposition}

\DeclareMathOperator*{\argmax}{arg\,max}
\allowdisplaybreaks

\begin{document}
% Page heads
%\markboth{G. Zhou et al.}{A Multifrequency MAC Specially Designed for WSN Applications}

% Title portion
\title{\Large The menu complexity of ``one-and-a-half-dimensional'' mechanism design} 
\author{Raghuvansh R. Saxena \thanks{Department of Computer Science, Princeton University, rrsaxena@cs.princeton.edu.} \and 
Ariel Schvartzman  \thanks{Department of Computer Science, Princeton University, acohenca@cs.princeton.edu.} \and
S. Matthew Weinberg \thanks{Department of Computer Science, Princeton University, smweinberg@princeton.edu.}}
\date{}
%\authorrunning{R. R. Saxena, A. Schvartzman and S. M. Weinberg} %mandatory. First: Use abbreviated first/middle names. Second (only in severe cases): Use first author plus 'et. al.'
%\Copyright{Raghuvansh Saxena, Ariel Schvartzman and Seth Matthew Weinberg}%mandatory, please use full first names. LIPIcs license is "CC-BY";  http://creativecommons.org/licenses/by/3.0/

%\subjclass{J.4 Social and Behavioral Sciences}% mandatory: Please choose ACM 1998 classifications from http://www.acm.org/about/class/ccs98-html . E.g., cite as "F.1.1 Models of Computation". 
%\keywords{Menu Complexity, 1.5-dimensional auctions, Fedex problem, approximate mechanism design}% mandatory: Please provide 1-5 keywords
%\EventEditors{??}
%\EventNoEds{??}
%\EventLongTitle{8th Innovations in Theoretical Computer Science (ITCS 2017)}
%\EventShortTitle{ITCS 2017}
%\EventAcronym{ITCS}
%\EventYear{2017}
%\EventDate{January 9--11, 2016}
%\EventLocation{Berkeley, California, USA}
%\EventLogo{}
%\SeriesVolume{??}
%\ArticleNo{??}

% NOTE! Affiliations placed here should be for the institution where the
%       BULK of the research was done. If the author has gone to a new
%       institution, before publication, the (above) affiliation should NOT be changed.
%       The authors 'current' address may be given in the "Author's addresses:" block (below).
%       So for example, Mr. Abdelzaher, the bulk of the research was done at UIUC, and he is
%       currently affiliated with NASA.

\maketitle
%\fancyfoot[R]{\footnotesize{\textbf{Copyright \textcopyright\ 2018 by SIAM\\
%Unauthorized reproduction of this article is prohibited}}}
\thispagestyle{empty}
\addtocounter{page}{-1}
\begin{abstract}
We study the menu complexity of optimal and approximately-optimal auctions in the context of the ``FedEx'' problem, a so-called ``one-and-a-half-dimensional'' setting where a single bidder has both a value and a deadline for receiving an item~\cite{Fiat:2016}. The menu complexity of an auction is equal to the number of distinct (allocation, price) pairs that a bidder might receive~\cite{HartN13}. We show the following when the bidder has $n$ possible deadlines:
\begin{itemize}[leftmargin=*]
\item \textbf{Exponential menu complexity is necessary to be exactly optimal:} There exist instances where the optimal mechanism has menu complexity $\geq 2^n-1$. This matches \emph{exactly} the upper bound provided by Fiat et al.'s algorithm, and resolves one of their open questions~\cite{Fiat:2016}.
\item \textbf{Fully polynomial menu complexity is necessary and sufficient for approximation:} For all instances, there exists a mechanism guaranteeing a multiplicative $(1-\epsilon)$-approximation to the optimal revenue with menu complexity $O(n^{3/2}\sqrt{\frac{\min\{n/\epsilon,\ln(v_{\max})\}}{\epsilon}}) = O(n^2/\epsilon)$, where $v_{\max}$ denotes the largest value in the support of integral distributions.
\item There exist instances where any mechanism guaranteeing a multiplicative $(1-O(1/n^2))$-approximation to the optimal revenue requires menu complexity $\Omega(n^2)$. 
\end{itemize}
Our main technique is the polygon approximation of concave functions~\cite{Rote91}, and our results here should be of independent interest. We further show how our techniques can be used to resolve an open question of~\cite{DevanurW17} on the menu complexity of optimal auctions for a budget-constrained buyer. 
\end{abstract}   
\newpage
\section{Introduction}
\label{sec:intro}

It is by now quite well understood that optimal mechanisms are far from simple: they may be randomized~\cite{Thanassoulis04, BriestCKW10, HartN13}, behave non-monotonically~\cite{HartR15, RubinsteinW15}, and be computationally hard to find~\cite{CaiDW13b, DaskalakisDT14, ChenDPSY14, Rubinstein16}. To cope with this, much recent attention has shifted to the design of simple, but approximately optimal mechanisms (e.g.~\cite{ChawlaHK07, ChawlaHMS10, HartN12, BabaioffILW14}). However, the majority of these works take a binary view on simplicity, developing simple mechanisms that guarantee constant-factor approximations. Only recently have researchers started to explore the tradeoff space between simplicity and optimality through the lens of \emph{menu complexity}.

Hart and Nisan first proposed the menu complexity as one quantitative measure of simplicity, which captures the \emph{number of different outcomes} that a buyer might see when participating in a mechanism~\cite{HartN13}. For example, the mechanism that offers only the grand bundle of all items at price $p$ (or nothing at price $0$) has menu complexity $1$. The mechanism that offers any single item at price $p$ (or nothing at price $0$) has menu complexity $n$, and randomized mechanisms could have infinite menu complexity. 

Still, all results to date regarding menu complexity have really been more qualitative than quantitative. For example, only just now is the state-of-the-art able to show that for a single additive bidder with independent values for multiple items and all $\varepsilon > 0$, the menu complexity required for a $(1-\varepsilon)$ approximation is finite~\cite{BabaioffGN17} (and even reaching this point was quite non-trivial). On the quantitative side, the best known positive results for a single additive or unit-demand bidder with independent item values require menu complexity exp(n) for a $(1-\varepsilon)$-approximation, but the best known lower bounds have yet to rule out that poly(n) menu complexity suffices for a $(1-\varepsilon)$-approximation in either case. In this context, our work provides \emph{the first nearly-tight quantitative bounds on menu complexity in any multi-dimensional setting}. 

\subsection{One-and-a-half dimensional mechanism design}
The setting we consider is the so-called ``FedEx Problem,'' first studied in~\cite{Fiat:2016}. Here, there is a single bidder with a value $v$ for the item and a deadline $i$ for receiving it, and the pair $(v,i)$ is drawn from an arbitrarily correlated distribution where the number of possible deadlines is finite ($n$). The buyer's value for receiving the item by her deadline is $v$, and her value for receiving the item after her deadline (or not at all) is $0$. While technically a two-dimensional problem, optimal mechanisms for the FedEx problem don't suffer the same undesirable properties as ``truly'' two-dimensional problems. Still, the space of optimal mechanisms is considerably richer than single-dimensional problems (hence the colloquial term ``one-and-a-half dimensional''). More specifically, while the optimal mechanism might be randomized, it has menu complexity at most $2^n-1$, and there is an inductive closed-form solution describing it. Additionally, there is a natural condition on each $F_i$ (the marginal distribution of $v$ conditioned on $i$) guaranteeing that the optimal mechanism is deterministic (and therefore has menu complexity $\leq n$).\footnote{This condition is called ``decreasing marginal revenues,'' and is satisfied by distributions with CDF $F$ and PDF $f$ such that $x\cdot f(x) - 1+F(x)$ is monotone non-decreasing.} 

A number of recent (and not-so-recent) works examine similar settings such as when the buyer has a value and budget~\cite{LaffontR96, CheG00, ChawlaMM11, DevanurW17}, or a value and a capacity~\cite{DevanurHP17}, and observe similar structure on the optimal mechanism. Such settings are quickly gaining interest within the algorithmic mechanism design community as they are rich enough for optimal mechanisms to be highly non-trivial, but not quite so chaotic as truly multi-dimensional settings.

\subsection{Our results}\label{sec:results}
We study the menu complexity of optimal and approximately optimal mechanisms for the FedEx problem. Our first result proves that the $2^n-1$ upper bound on the menu complexity of the optimal mechanism provided by Fiat et al.'s algorithm is \emph{exactly} tight:

\begin{theorem}\label{thm:exact}
For all $n$, there exist instances of the FedEx problem on $n$ days where the menu complexity of the optimal mechanism is $2^n-1$. 
\end{theorem}

From here, we turn to approximation and prove our main results. First, we show that fully polynomial menu complexity suffices for a $(1-\varepsilon)$-approximation. The guarantee below is always $O(n^2/\varepsilon)$, but is often improved for specific instances. Below, if the FedEx instance happens to have integral support and the largest value is $v_{\max}$, we can get an improved bound (but if the support is continuous or otherwise non-integral, we can just take the $n/\varepsilon$ term instead).\footnote{Actually our bounds can be be improved to replace $v_{\max}$ with many other quantities that are always $\leq v_{\max}$, and will still be well-defined for continuous distributions, more on this in Section~\ref{sec:UB}.}

\begin{theorem}\label{thm:ubmain}
For all instances of the FedEx problem on $n$ days, there exists a mechanism of menu complexity $O\left(n\sqrt{\frac{\min\{n/\varepsilon,\ln(v_{\max})\}}{\varepsilon/n}}\right)$ guaranteeing a $(1-\varepsilon)$ approximation to the optimal revenue. 
\end{theorem}

In Theorem~\ref{thm:ubmain}, observe that for any fixed instance, as $\varepsilon \rightarrow 0$, our bound grows like $O(1/\sqrt{\varepsilon})$ (because eventually $n/\varepsilon$ will exceed $\ln(v_{\max})$). Similarly, our bound is always $O(n^2/\varepsilon)$ for any $v_{\max}$. Both of these dependencies are provably tight for our approach (discussed shortly in Section~\ref{sec:techniques}), and in general tight up to a factor of $\sqrt{n\log n}$.\footnote{The gap of $\sqrt{n\log n}$ comes as our upper bound approach requires that we lose at most $\varepsilon \mathsf{OPT}/n$ ``per day,'' while our lower bound approach shows that any mechanism with lower menu complexity loses at least $\varepsilon \mathsf{OPT}$ on some day.}

\begin{theorem}\label{thm:lbmain}
For all $n$, there exists an instance of the FedEx problem on $n$ days with $v_{\max} = O(n)$, such that the menu complexity of every $(1-O(1/n^2))$-optimal mechanism is $\Omega(n^2)$. 
\end{theorem}

We consider Theorems~\ref{thm:ubmain} and~\ref{thm:lbmain} to be our main results, with Theorem~\ref{thm:exact} motivating the study of approximation in the first place. Taken together, the picture provided by these results is the following:
\begin{itemize}
\item Exactly optimal mechanisms can require exponential menu complexity (Theorem~\ref{thm:exact}), while $(1-\varepsilon)$-approximate mechanisms exist with fully polynomial menu complexity (Theorem~\ref{thm:ubmain}).
\item The menu complexity required to guarantee a $(1-\varepsilon)$-approximation is nailed down within a multiplicative $\sqrt{n\log n}$ gap, and lies in $\left[\Omega\left(\sqrt{n/\log n} \cdot\sqrt{\frac{\min\{n/\varepsilon,\ln (v_{\max})\}}{\varepsilon / n}}\right), O\left(n \cdot\sqrt{\frac{\min\{n/\varepsilon,\ln (v_{\max})\}}{\varepsilon / n}}\right)\right]$ (lower bound: Theorem~\ref{thm:lbmain}, upper bound Theorem~\ref{thm:ubmain}). 
\end{itemize}

\subsection{Our techniques}\label{sec:techniques}
We'll provide an intuitive proof overview for each result in the corresponding technical section, but we briefly want to highlight one aspect of our approach that should be of independent interest. 

It turns out that the problem of revenue maximization with bounded menu complexity really boils down to a question of how well piece-wise linear functions with bounded number of segments can approximate concave functions (we won't get into details of \emph{why} this is the case until Section~\ref{sec:UB}). This is a quite well-studied problem called \emph{polygon approximation} (e.g.~\cite{Rote91,YangG97,Burkard91}). Questions asked here are typically of the form ``for a concave function $f$ and interval $[0,v_{\max}]$ such that $f'(0) = 1$, $f'(v_{\max}) = 0$, what is the minimum number of segments a piece-wise linear function $g$ must have to guarantee $f(x) \geq g(x) \geq f(x) - \varepsilon$ for all $x \in [0,v_{\max}]$?''

The answer to the above question is $\Theta(\sqrt{v_{\max}/\varepsilon})$~\cite{Rote91, YangG97}. This bound certainly suffices for our purposes to get \emph{some} bound on the menu complexity of $(1-\varepsilon)$-approximate auctions, but it would be much weaker than what Theorem~\ref{thm:ubmain} provides (we'd have linear instead of logarithmic dependence on $v_{\max}$, and no option to remove $v_{\max}$ from the picture completely). Interestingly though, for our application absolute additive error doesn't tightly characterize what we need (again, we won't get into why this is the case until Section~\ref{sec:UB}). Instead, we are really looking for the following kind of guarantee, which is a bit of a hybrid between additive and multiplicative: for a concave function $f$ and interval $[0,v_{\max}]$ such that $f'(0) = 1$, $f'(v_{\max}) = 0$, what is the minimum number of segments a piece-wise linear function $g$ must have to guarantee $f(x) \geq g(x) \geq f(x) - \varepsilon - \varepsilon (f(v_{\max})-f(0))$?

At first glance it seems like this really shouldn't change the problem at all: why don't we just redefine $\varepsilon':=\varepsilon(1+f(v_{\max})-f(0))$ and plug into upper bounds of Rote for $\varepsilon'$? This is indeed completely valid, and we could again chase through and obtain some weaker version of Theorem~\ref{thm:ubmain} that also references additional parameters in unintuitive ways. But it turns out that for all examples in which this $\Omega(\sqrt{v_{\max}/\varepsilon})$ dependence is tight, there is actually quite a large gap between $f(0)$ and $f(v_{\max})$, and a greatly improved bound is possible (which replaces the linear dependence on $v_{\max}$ with logarithmic dependence, and provides an option to remove $v_{\max}$ from the picture completely at the cost of worse dependence on $\varepsilon$).

\begin{theorem}\label{thm:polygon}
For any concave function $f$ and any $\varepsilon > 0$ such that $f'(0) \leq 1$, $f'(v_{\max}) \geq 0$, there exists a piece-wise linear function $g$ such that $f(x) \geq g(x) \geq f(x) - \varepsilon (1+f(v_{\max})-f(0))$ with $\Theta(\sqrt{\ln(v_{\max})/\varepsilon})$ segments, and this is tight. 

If one wishes to remove the dependence on $v_{\max}$, then one can replace the bound with $\Theta(1/\varepsilon)$, which is also tight (among bounds that don't depend on $v_{\max}$). 
\end{theorem}

The proof of Theorem~\ref{thm:polygon} is self-contained and appears in Section~\ref{sec:UB}. Both the statement of Theorem~\ref{thm:polygon} and our proof will be useful for future work on menu complexity, and possibly outside of mechanism design as well - to the best of our knowledge these kinds of hybrid guarantees haven't been previously considered.\footnote{Interestingly (and completely unrelated to this work), hybrid additive-multiplicative approximations for core problems in online learning have also found use in other recent directions in AGT~\cite{FosterLLST16, BubeckDHN17}.}

\subsection{Related work}\label{sec:related}
\textbf{Menu complexity.} Initial results on menu complexity prove that for a single additive or unit-demand bidder with arbitrarily correlated item values over just $2$ items, there exist instances where the optimal (randomized, with infinite menu complexity) mechanism achieves infinite revenue, while any mechanism of menu complexity $C$ achieves revenue $\leq C$ (so no finite approximation is possible with bounded menu complexity)~\cite{BriestCKW10, HartN13}. This motivated follow-up work subject to assumptions on the distributions, such as a generalized hazard rate condition~\cite{TangW14}, or independence across item values~\cite{DaskalakisDT13, BabaioffGN17}. Even for a single bidder with independent values for two items, the optimal mechanism could have uncountable menu complexity~\cite{DaskalakisDT13}, motivating the study of approximately optimal mechanisms subject to these assumptions. Only just recently did we learn that the menu complexity is indeed finite for this setting~\cite{BabaioffGN17}. 

%It is also worth noting that a related notion of ``additive'' menu complexity has also been considered. For the FedEx problem (or unit-demand bidders in general), menu complexity and additive menu complexity are identical, so our results apply to both notions. Without getting into a formal definition, the difference for additive bidders is the following: technically, the mechanism that offers any subset $S$ of items at price $\sum_{i \in S} p_i$ has menu complexity $2^n-1$. But really the menu has a very structured description of size $n$, so we'd like a complexity measure that captures this (and additive menu complexity exactly captures this). Babaioff et al. do rule out the possibility that a fully polynomial menu complexity suffices for a $(1-\varepsilon)$-approximation to the optimal revenue, but in their example the optimal mechanism has additive menu complexity $n$ (and in fact just sells each item separately as described above)~\cite{BabaioffGN17}. 

It is also worth noting that other notions of simplicity have been previously considered as well, such as the sample complexity (how many samples from a distribution are required to learn an approximately optimal auction?). Here, quantitative bounds are known for the single-item setting (where the menu complexity question is trivial: optimal mechanisms have menu complexity $1$)~\cite{ColeR14, HuangMR15, DevanurHP16, GonczarowskiN17}, but again only binary bounds are known for the multi-item setting: few samples suffice for a constant-factor approximation if values are independent~\cite{MorgensternR15, MorgensternR16}, while exponentially many samples are required when values are arbitrarily correlated~\cite{DughmiHN14}. In comparison to works of the previous paragraphs, we are the first to nail down ``the right'' quantitative menu complexity bounds in any multi-dimensional setting. 

\textbf{One-and-a-half dimensional mechanism design.} One-and-a-half dimensional settings have been studied for decades by economists, the most notable example possibly being that of a single buyer with a value and a budget~\cite{LaffontR96, CheG00}. Recently, such problems have become popular within the AGT community as optimal auctions are more involved than single-dimensional settings, but not quite so chaotic as truly multidimensional settings~\cite{Fiat:2016, DevanurW17, DevanurHP17}. Each of these works focus exclusively on exactly optimal mechanisms (and exclusively on positive results). In comparison, our work is both the first to prove lower bounds on the complexity of (approximately) optimal mechanisms in these settings, and the first to provide nearly-optimal mechanisms that are considerably less complex. 

\textbf{Polygon approximation.} Prior work on polygon approximation is vast, and includes, for instance, core results on univariate concave functions~\cite{Rote91, Burkard91, YangG97}, the study of multi-variate functions~\cite{Bronstein08, GlasauerG09, DaskalakisDY16}, and even applications in robotics~\cite{Boissonnat07}. The more recent work has mostly been pushing toward better guarantees for higher dimensional functions. To the best of our knowledge, the kinds of guarantees we target via Theorem~\ref{thm:polygon} haven't been previously considered, and could prove more useful than absolute additive guarantees for some applications.

\subsection{Organization}
In Section~\ref{sec:definitions}, we formally describe the FedEx problem and recap the main result of~\cite{Fiat:2016}. In Section~\ref{sec:LB} we present an instance of the FedEx problem whose menu complexity for optimal auctions is exponential, the worst possible. In Section~\ref{sec:UB} we present a mechanism that guarantees a (1-$\varepsilon$) fraction of the optimal revenue with a menu complexity of $O(\frac{n^2}{\varepsilon})$. We also explain the connection between approximate auctions and polygon approximation. In Section~\ref{sec:LBA} we present an instance of the FedEx problem that requires a menu complexity of $\Omega(n^2)$ in order to approximate the revenue within $1-O(1/n^2)$. In Section~\ref{sec:reg3daysplit} we use similar techniques to those of Section~\ref{sec:LB} to construct an example resolving an open question of~\cite{DevanurW17}.\footnote{Specifically,~\cite{DevanurW17} ask whether the optimal mechanism for a single buyer with a private budget and a regular value distribution conditioned on each possible budget is deterministic. The answer is yes if we replace ``regular'' with ``decreasing marginal revenues,'' or ``private budget'' with ``public budget.'' We show that the answer is no in general: the optimal mechanism, even subject to regularity, could be randomized.}

\section{Preliminaries}
\label{sec:definitions}
We consider a single bidder who's type depends on two parameters: a value $v$ and a deadline $i \in [n]$. Deterministic outcomes that the seller can award are just a day $\in [n]$ to ship the item, or to not ship the item at all (and the seller may also randomize over these outcomes). A buyer receives value $v$ if the item is shipped by her deadline, and $0$ if it is shipped after her deadline (or not at all). 

%Thus the bidder reports a pair $(v,i)$ and the seller returns a price and shipping date. The bidder gets no utility from the item if it is shipped after the true deadline $i$. This problem was first considered in~\cite{Fiat:2016} so we will follow their notation for the most part.  

The types $(v,i)$ are drawn from a known (possibly correlated) distribution $\mathcal{F}$. Let $q_i$ denote the probability that the bidder's deadline is $i$ and $\mathcal{F}_i$ the marginal distribution of $v$ conditioned on a deadline of $i$. For simplicity of exposition, in several parts of this paper we'll assume that $\mathcal{F}$ is supported on $\{0,1,\ldots,v_{\max}\} \times \{1,\ldots, n\}$. This assumption is w.l.o.g., and all results extend to continuous distributions, or distributions with arbitrary discrete support if desired~\cite{CaiDW16s}. 

%We will denote cumulative functions with capital letters, and density functions with lower case letters. Let $\pi(v,i), p(v,i)$ denote the probability that the package is shipped on day $i$ and the expected price the buyer incurs when $(v,i)$ is reported to the mechanism. Throughout this paper we will  work with discrete distributions for simplicity, but the results presented can be extended to the continuous domain (see~\cite{CaiDW16s} for a discussion). Thus we assume that $\mathcal{F}$ is supported on $\{0, 1, ... , v_{\max} \} \times \{1, ..., n \}$. 

%We will be interested in optimal and approximately optimal auctions in this setting. By the revelation principle, for the former, we can restrict our attention to incentive-compatible auctions.  Misreporting later deadlines (i.e. $i' > i$) is never helpful to the bidder since they get no utility for receiving the item after day $i$ and will have to incur in a possibly non-zero price. By a standard argument, we do not need to consider all possible misreports, only those that are adjacent to type $(v,i)$~\cite{DW17, Fiat:2016}. 

In Appendix~\ref{app:preliminaries}, we provide the standard linear program whose solution yields the revenue-optimal auction for the FedEx problem. We only note here the relevant incentive compatibility constraints (observed in~\cite{Fiat:2016}). First, note that w.l.o.g. whenever the buyer has deadline $i$, the optimal mechanism can ship her the item (if at all) exactly on day $i$. Shipping the item earlier doesn't make her any happier, but might make the buyer interested in misreporting and claiming a deadline of $i$ if her deadline is in fact earlier. Next note that, subject to this, the buyer never has an incentive to overreport her deadline, but she still might have incentive to underreport her deadline (or misreport her value). 

We will be interested in understanding the \emph{menu complexity} of auctions, which is the number of different outcomes that, depending on the buyer's type, are ever selected. If $\pi(v,i)$ denotes the probability that a buyer with value $v$ and deadline $i$ receives the item, then we define the $i$-deadline menu complexity to be the number of distinct options on deadline $i$ ($|\{p | \exists v, \pi(v, i) = p\}|$). The menu complexity then just sums the $i$-deadline menu complexities, and we will sometimes refer also to the ``deadline menu complexity'' as the maximum of the $i$-deadline menu complexities.

%\begin{definition}
%Consider an incentive compatible and individually rational mechanism. By the taxation principle, we can just replace the auction by a menu of options $M = (x, p)$ where $x$ is the probability the buyer gets the item and pays $p$, and where the buyer chooses an option that maximizes expected utility. The menu complexity of an auction is the number of entries of such a menu $M$. 	
%\end{definition}
	
%Due to the one-and-a-half dimensional nature of the FedEx problem we take an additional step and define the \emph{$i$-deadline menu complexity} as the number of options offered conditioned on day $i$, and define the \emph{deadline menu complexity} as the maximum deadline complexity over all deadlines.  

%\begin{definition}[Canonical virtual value function] 
%Given a probability mass function $f: [v_{\max}] \rightarrow [0,1]$, we define the canonical virtual value function of $f$ as the function $\phi_f: [v_{\max}] \rightarrow \mathbb{R}$ such that
%$$\phi(v) =  v - \frac{1-F(v)}{f(v)}$$ 
%where $F$ is the cumulative mass function of $f$.
%\end{definition}

%\begin{lemma}
%\label{lem:virtualrev}
%Let $f$ be a distribution over $[v_{\max}]$, and let $\phi_f, R_f$ be the corresponding virtual value and revenue curve, respectively. Then, for all $v \in [v_{max}-1]$ 
%$$R_f(v) - R_f(v-1) = \phi_f(v) f(v) .$$ 
%\end{lemma}

%\begin{proof}
%This follows from some simple calculations: 

%\begin{equation*}
%\begin{aligned}
%R_f(v) - R_f (v-1) & = v(1-F(v-1))-(v+1)(1-F(v)) \\
%& = v(F(v)-F(v-1)) - (1-F(v)) \\
%& = vf(v) - (1-F(v)) =  f(v) \phi_f(v) .
%\end{aligned}
%\end{equation*}
%\end{proof}

\subsection{Optimal auctions for the FedEx problem} 
Here, we recall some tools from~\cite{Fiat:2016} regarding optimal mechanisms for the FedEx problem. The first tool they use is the notion of a \emph{revenue curve}.\footnote{For those familiar with revenue curves, note that this revenue curve is intentionally drawn in value space, and not quantile space.}

\begin{definition}[Revenue curves] 
For a given deadline $i$, define the $i^{th}$ revenue curve $R_i$ so that
$$R_i(v) = q_i \cdot v \cdot \Pr_{x \leftarrow \mathcal{F}_i}[x \geq v].$$
\end{definition}

Intuitively, $R_i(v)$ captures the achievable revenue by setting price $v$ exclusively for consumers on deadline $i$. It is also necessary to consider the \emph{ironed} revenue curve, defined below.

\begin{definition} [Ironed revenue curves] For any revenue curve $R_i$, define $\tilde{R}_i$ to be its upper concave envelope.\footnote{That is, $\tilde{R}_i$ is the smallest concave function such that $\tilde{R}_i(x) \geq R_i(x)$ for all $x$.} We say $\tilde{R}_i$ is ironed at $v$ if $\tilde{R}_i(v) \neq R_i (v)$, and we call $[x,y]$ an ironed interval of $\tilde{R}_i$ if $\tilde{R}_i$ is not ironed at $x$ or $y$, but is ironed at $v$ for all $v \in (x,y)$.\end{definition}

Of course, it is not sufficient to consider each possible deadline of the buyer in isolation. In particular, offering certain options on day $i$ constrains what can be offered on days $\geq i$ subject to incentive compatibility. For instance, if some $(v,i)$ pair receives the item with probability $1$ on day $1$ for price $p$, no bidder with a deadline $\geq 1$ will ever choose to pay $>p$. So we would also like a revenue curve that captures the optimal revenue we can make from days $\geq i$ conditioned on selling the item deterministically at price $p$ on day $i$. It's not obvious how to construct such a curve, but this is one of the main contributions of~\cite{Fiat:2016}, stated below.

\begin{definition} \label{def:revcurves}

Let $R_{\geq n} (v) := R_n (v),$ and $r_{\geq n} := \argmax_v R_{\geq n}(v).$ Define for $i = n-1$ to $1$:

\[  R_{\geq i}(v) = \left\{ 
\begin{array}{ll}
      R_{i}(v) + \tilde{R}_{\geq i+1} (v) & v < r_{\geq i+1} \\
      R_{i}(v) + \tilde{R}_{\geq i+1} (r_{\geq i+1}) & v \geq r_{\geq i+1}. \\
\end{array} 
\right. \]
\end{definition}
\begin{lemma}[\cite{Fiat:2016}]\label{lem:geqi}
$R_{\geq i}(v)$ is the optimal revenue of any mechanism that satisfy the following:
\begin{itemize} 
\item The buyer can either receive the item on day $i$ and pay $v$, or receive nothing/pay nothing.
\item The buyer cannot receive the item on any day $< i$.
\end{itemize}
Moreover, for any $v_1 < \ldots < v_k$, and $a_i(1),\ldots,a_i(k) \geq 0$ such that $\sum_j a_i(j) \leq 1$, $\sum_j a_i(j)R_{\geq i}(v_j)$ is the optimal revenue of any mechanism that satisfy the following:
\begin{itemize}
\item The buyer can receive the item on day $i$ with probability $\sum_{j \leq \ell} a_i(j)$ and pay $\sum_{j \leq \ell} a_i(j) v_j$, for any $\ell \in [k]$ (or not receive the item on day $i$ and pay nothing). 
\item The buyer cannot receive the item on any day $< i$.
\end{itemize}
\end{lemma}

Finally, we describe the optimal mechanism provided by~\cite{Fiat:2016}, which essentially places mass optimally upon each day's revenue curve, subject to constraints imposed by the decisions of previous days (a more detailed description appears in Appendix~\ref{app:preliminaries}, but the description below will suffice for our paper). First, simply set any price $p$ maximizing $R_{\geq 1}(p)$ to receive the item on day $1$ (as day $1$ is unconstrained by previous days). Now inductively, assume that the options for day $i$ have been set and we're deciding what to do for day $i+1$. If the menu options offered on day $i$ are $(\pi_0,p_0),\ldots, (\pi_k, p_k)$ (interpret the option $(\pi_j,p_j)$ as ``charge $p_j$ to ship the item on day $i$ with probability $\pi_j$''), think of this instead as a distribution over prices, where price $\frac{p_j-p_{j-1}}{\pi_j-\pi_{j-1}}$ has mass $\pi_j-\pi_{j-1}$.\footnote{This is the standard transformation between ``lotteries'' and ``distributions over prices'' (e.g.~\cite{RileyZ83}).} For each such price $p$, it will undergo one of the following three operations to become an option for day $i+1$.
\begin{itemize}
\item If $p \geq r_{\geq i+1}$, move all mass from $p$ to $r_{\geq i+1}$.
\item If $\hat{R}_{\geq i+1}$ is not ironed at $p$, and $p \leq r_{\geq i+1}$, keep all mass at $p$.
\item If $\hat{R}_{\geq i+1}$ is ironed at $p$, and $p \leq r_{\geq i+1}$, let $[x,y]$ denote the ironed interval containing $p$, and let $qx + (1-q)y = p$. Move a $q$ fraction of the mass at $p$ to $x$, and a $(1-q)$ fraction of the mass at $p$ to $y$.
\end{itemize}

Once the mass is resettled, if there is mass $a_i(j)$ on price $p_j$ for $p_1 < \ldots < p_k$, the buyer will have the option to receive the item on day $i$ with probability $\sum_{j \leq \ell} a_i(j) $ for price $\sum_{j \leq \ell} a_i(j) p_j$ for any $\ell \in [k]$ (or not at all). Note that due to case three in the transformation above, there could be up to twice as many menu options on day $i$ as day $i-1$. 

\begin{theorem}[\cite{Fiat:2016}]
The allocation rule described above is the revenue-optimal auction.
\end{theorem}

\section{Optimal Mechanisms Require Exponential Menu Complexity}
\label{sec:LB}
In this section we overview our construction for an instance of the FedEx problem with $v_{\max}$ integral values for each day and $n \leq \log(v_{\max})$ days where the $i$-deadline menu complexity of the optimal mechanism is $2^{i-1}$ for all $i$ (and this is the maximum possible~\cite{Fiat:2016}), implying that the menu complexity is $2^n - 1$. Note that the deadline menu complexity is always upper bounded by $v_{\max}$, so $v_{\max}$ must be at least $2^n$.

At a high level, constructing the example appears straight-forward, once one understands Fiat et al.'s algorithm (end of Section~\ref{sec:definitions}). Every menu option from day $i$ is either ``shifted'' to $r_{\geq i+1}$, ``copied,'' or ``split.'' If the option is shifted or copied, it spawns only a single menu option on day $i+1$, while if split it spawns two (hence the upper bound of $2^n-1$). So the goal is just to construct an instance where every option is split on every day.

Unfortunately, this is not quite so straight-forward: whether or not an option is split depends on whether it lies inside an ironed interval in this $R_{\geq i}$ curve, which is itself the sum of revenue curves (some ironed and some not), and going back and forth between distributions and sums of revenue curves is somewhat of a mess. So really what we'd like to do is construct the $R_{\geq i}$ curves directly, and be able to claim that there exists a FedEx input inducing them. While not every profile $(R_{\geq 1},\ldots,R_{\geq n})$ of curves is valid, we do provide a broad class of curves for which it is somewhat clean to show that there exists a FedEx input inducing them.

From here, it is then a matter of ensuring that we can find the revenue curve profiles we want (where for every day $i$, every menu option is split, because it is inside an ironed interval in $R_{\geq i}$) within our class. We'll highlight parts of our construction below, but most details are in Appendix~\ref{sec:optworstcase}.
\begin{lemma}
\label{optrevexp}
For any $v_{\max}$ and $n \leq \log(v_{\max})$, there exists an input to the FedEx problem such that: 
\begin{itemize}
	\item $R_1$ is maximized at $v_{\max}/2$ (that is, $R_1(v_{\max}/2) \geq R_1(x) \ \forall x$) and has no ironed intervals. 
	\item For all $i>1$, $\tilde{R}_i$ has a maximizer at price $v \geq 2^{i}(2^{n-i}-1)$ and has ironed intervals $[2^{n-i}+k2^{n-i+2}, 2^{n-i} + k2^{n-i+2}+2^{n-i+1}]$ for $k \in \{0,\ldots,2^{i-2}-1\}$.
	\item $\tilde{R}_{i}$ (the ironed revenue curve) is a constant function for all $i \geq 2$.\footnote{Note that it is possible for two disjoint ironed intervals to have the same slope.} 
	\item $R_{\geq i}$ has the same ironed intervals as ${R}_{i}$. In fact, $\forall x$, $R_{\geq i}(x) = R_{i}(x)+c$ for some constant $c$.
\end{itemize}
\end{lemma} 

We include in Figure~\ref{fig1:fourdayrev} a picture of the generated revenue curves for $n = 4$. As a result of this construction, we see that $R_{\geq i}$ has $2^{i-2}$ ironed intervals, whose endpoints themselves lie in ironed intervals of $R_{\geq i+1}$. This guarantees that all menu options from day $i$ (which are guaranteed to be endpoints of ironed intervals) are split into two options on day $i+1$. The proof of Theorem~\ref{thm:expmenu} (which implies Theorem~\ref{thm:exact}) formalizes this.

\begin{theorem}
\label{thm:expmenu}
The optimal mechanism for any instance satisfying the conditons of Lemma~\ref{optrevexp} has $i$-deadline complexity $2^{i-1}$ for all $i$, and menu complexity $2^n -1 $. 
\end{theorem}

The proof of Theorem~\ref{thm:expmenu} is presented at the end of Section~\ref{sec:optworstcase} of the Appendix.

\section{Approximately Optimal Mechanisms with Small Menus}
\label{sec:UB}
In this section, we describe a mechanism that attains at least $1-\varepsilon$ fraction of the optimal revenue for any FedEx instance with menu complexity $O\left(n\sqrt{\frac{n}{\varepsilon}\min\left(\frac{n}{\varepsilon}, \log v_{\max}\right)}\right)$, which proves \autoref{thm:ubmain}. Most proofs appear in Appendix~\ref{app:UB}, but we overview our approach here.

%In this section, we'll prove \autoref{thm:ubmain}. We start by drawing a connection between menu complexity and polygon approximation. \rrsnote{Haven't defined polygon approximation yet}

Our main approach is to use the polygon approximation of concave functions applied to revenue curves. For a sequence of points $X$ in the domain of a function $f$, the polygon approximation $\tilde{f}_X$ of a function with respect to $X$ is the piecewise linear function formed by connecting the points $(x,f(x))$ for $x \in X$ by line segments. Thus, if the sequence $X$ has $n$ points, the function $\tilde{f}_X$ will have $n-1$ segments.  For a concave function $f$, the line joining $(x_1, f(x_1))$ and $(x_2, f(x_2))$ for any two points $x_1$ and $x_2$, lies entirely below the function $f$. Thus, for concave functions $f$, we have for any sequence $X$, the value of $f(x) - \tilde{f}_X(x) \geq 0$. Typically, for a `good' polygon approximation, one requires for $\varepsilon > 0$, that $f(x) - \varepsilon \leq \tilde{f}_X(x) \leq f(x)$. 

It turns out that the question of approximating revenue with low menu complexity boils down to a question of approximating \emph{revenue curves} with piecewise-linear functions of few segments. The connection isn't quite obvious, but isn't overly complicated. Without getting into formal details, here is a rough sketch of what's going on:
\begin{itemize}
\item Recall the Fiat et al. procedure to build the optimal mechanism: menu options from deadline $i-1$ might be ``split'' into two options for deadline $i$ if they lie inside an ironed interval of $\tilde{R}_{\geq i}$. This might cause the menu complexity to double from one deadline to the next.
\item Instead, we want to create at most $k$ ``anchoring points'' on each revenue curve. For a menu option from deadline $i-1$, instead of distributing it to the endpoints of its ironed interval, we distribute it to the two nearest anchor points.
\item By~\autoref{lem:geqi}, we know exactly how to evaluate the revenue lost by this change, and it turns out this is captured by the maximum gap between $\tilde{R}_{\geq i}(\cdot)$ and the polygon approximation obtained to $\tilde{R}_{\geq i}(\cdot)$ (this isn't obvious, but not hard. See Appendix~\ref{app:UB}). 
\item Finally, it turns out that the $i$-deadline menu complexity with at most $k$ anchoring points is at most $2k$ (also not quite obvious, but also not hard). So the game is to find few anchoring points that obtain a good polygon approximation to each revenue curve. \autoref{cor:curves} of \autoref{prop:curves} describes the reduction formally, but all related proofs are in \autoref{app:UB}.
\end{itemize}

\begin{corollary}\label{cor:curves}
Consider a FedEx instance with $n$ deadlines. For all $i \in \{1,2,\cdots,n\}$, let $g_i$ be the function $\tilde{R}_{\geq i}$ defined in \autoref{def:revcurves}, and let $X_i$ be a sequence of $k_i$ points in $[0,r_{\geq i}]$ such that for all $x \leq r_{\geq i}$, we have $g_i(x) - \varepsilon_i \leq  \tilde{g_i}_{X_i}(x) \leq  g_i(x) $. Then there exists a mechanism with $i$-deadline menu complexity $2k_i$ (and menu complexity $2 \sum_i k_i$) whose revenue is at least $ \mathsf{OPT} - \sum_{i=1}^n\varepsilon_i$.

Here, $\mathsf{OPT}$ denotes the optimal revenue of the FedEx instance.
\end{corollary}

At this point, it seems like the right approach is to just set each $\varepsilon_i = \varepsilon \cdot \mathsf{OPT}/n$ and plug into the best existing bounds on polygon approximation. In some sense this is correct, but the menu complexity bounds one would obtain are far from optimal. The main insight is that we know something about the curves we wish to approximate: $\tilde{R}(x) \leq \mathsf{OPT}$ for all $x$, and we want to leverage this fact if it can give us better guarantees. Additionally, if all values are integral in the range $\{1,\ldots, v_{\max}\}$, we wish to leverage this fact as well, as it implies that an additive $\varepsilon$ loss is also OK, as $\mathsf{OPT} \geq 1$. It turns out that both facts can indeed be leveraged to obtain much stronger approximation guarantees than what are already known (essentially replacing $v_{\max}$ with $\ln (v_{\max})$ in previous bounds), stated in~\autoref{thm:polygonapproximation} below.

\begin{theorem} \label{thm:polygonapproximation}  For any $\varepsilon >0$ and concave function $f:[0,v_{\max}] \to [0,\infty)$ such that $f(0) = 0$, $f^+(0) \leq 1$, $f^-(v_{\max}) \geq 0$\footnote{We use $f^+$ to denote the right hand derivative and $f^-$ to denote the left hand derivative.}, there exists a sequence $X$ of at most $O\left(\min\left\{1/\varepsilon,\sqrt{\frac{\log v_{\max}}{\varepsilon}}\right\}\right)$ points such that for all $x \in [0,v_{\max}],$
$$f(x)-\varepsilon \left(1+f(v_{\max})\right) \leq \tilde{f}_X(x) \leq f(x).$$
\end{theorem}

The proof of~\autoref{thm:ubmain} follows from \autoref{cor:curves} and~\autoref{thm:polygonapproximation} together with a little bit of algebra, and is deferred until Appendix~\ref{app:UB}.

Finally, we remark on some alternative terms that can be taken to replace $v_{\max}$ in~\autoref{thm:ubmain}. It will become clear why these replacements are valid after reading the proof of~\autoref{thm:ubmain}, but we will not further justify the validity of these replacements here.
\begin{itemize}
\item First, for instances with integral valuations, we may replace $v_{\max}$ everywhere with $\max_i r_{\geq i}$. This is essentially because we don't actually need to approximate $\tilde{R}_{\geq i}$ on the entire interval $[0,v_{\max}]$, but only the interval $[0,r_{\geq i}]$. 
\item We may further define $q = \max_i r_{\geq i}/\mathsf{OPT}$ for any (not necessarily integral, possibly continuous) instance, and replace $v_{\max}$ everywhere with $q$, even for non-integral instances. This is essentially because we only used the integrality assumption to guarantee that $\mathsf{OPT} \geq 1$.
\item Finally, if $p_{\geq i}$ denotes the probability that the buyer has value at least $r_{\geq i}$ and deadline at least $i$, observe that $\mathsf{OPT} \geq r_{\geq i}\cdot p_{\geq i}$. So if the probability of sale at each $r_{\geq i}$ is at least $p$, we may observe that $q \geq 1/p$ (where $q$ is defined as in the previous bullet) and replace $v_{\max}$ with $1/p$ everywhere. 
\end{itemize}

The bullets above suggest that the ``hard'' instances (where some instance-specific parameter shows up in order to maintain optimal dependence on $\varepsilon$) are those where most of the revenue comes from very infrequent events where the buyer has an unusually high value. Due to the intricate interaction between different deadlines, these parameters can't be circumvented with simple discretization arguments, or by improved polygon approximations (provably, see Section~\ref{sec:tightpolygon}), but it is certainly interesting to see if other arguments might allow one to replace $\log v_{\max}$ with (for example) something like $\log (n/\varepsilon)$.

\subsection{A tight example for polygon approximation} \label{sec:tightpolygon}
It turns out that the guarantees provided by~\autoref{thm:polygonapproximation} are tight. Specifically, if no dependence on $v_{\max}$ is desired, then $1/\varepsilon$ is the best  bound achievable. Also, if it's acceptable to depend on both $v_{\max}$ and $\varepsilon$, then the bound of $\sqrt{\frac{\log v_{\max}}{\varepsilon}}$ in \autoref{thm:polygonapproximation} is tight. Taken together, this means that $O\left(\min\left\{1/\varepsilon,\sqrt{\frac{\log v_{\max}}{\varepsilon}}\right\}\right)$ lies at the Pareto frontier of the dependences achievable as a function of both $v_{\max}$ and $\varepsilon$.  The examples proving tightness of these bounds are actually quite simple, and \emph{provably the worst possible examples} (proof of the below claim appears in Appendix~\ref{app:UB})

\begin{proposition}\label{prop:worstexample}
Let $f$ be a concave function on $[0,v_{\max}]$, and let there be no polygon approximation of $f$ using $k$ segments for additive error $\varepsilon$. Then there exists a concave function $g$ over $[0,v_{\max}]$ satisfying:
\begin{itemize}
\item There is no polygon approximation of $g$ using $k$ segments for additive error $\varepsilon$.
\item $f(0) = g(0)$, $f^+(0) = g^+(0)$, $f(v_{\max}) = g(v_{\max})$, $f^-(v_{\max}) = g^-(v_{\max})$. 
\item $g$ is piecewise-linear with $2k$ segments.
\end{itemize}
\end{proposition}

\section{Tightness of the approximation scheme}
\label{sec:LBA}
Finally, we construct an instance of the FedEx problem that is hard to approximate with small menu complexity. We try to reason similar to the example constructed in Section~\ref{sec:LB}, but things are trickier here. In particular, the challenge in Section~\ref{sec:LB} was in mapping between distributions and revenue curves. But once we had the revenue curves, it was relatively straight-forward to plug through Fiat et al.'s algorithm \cite{Fiat:2016} and ensure that the optimal auction had high menu complexity.

Already nailing down the behavior of an optimal auction was tricky enough, but we now have to consider \emph{every} approximately optimal auction (almost all of which don't necessarily result from Fiat et al.'s algorithm (see, e.g. Section~\ref{sec:UB})). Indeed, one can imagine doing all sorts of strange things on any day $i$ that are suboptimal, but might somehow avoid the gradual buildup in the $i$-deadline menu complexity.\footnote{For example, an $\varepsilon$-approximate menu could set price $0$ or $\infty$ with probability $\varepsilon$ for shipment on any day, or something much more chaotic.}

To cope with this, our approach has two phases: first, we characterize a restricted class of auctions that we call \emph{clean}. At a very high level, clean auctions never make ``bizarre'' choices on day $i$ that both decrease the revenue gained on day $i$ and \emph{strictly} increase constraints on choices available for future days. To have an example in mind: if the revenue on day $1$ is maximized by setting a price of $1$, it \emph{might} make sense to set price $2$ to receive the item on day $1$ instead, as this relaxes constraints on future days, and maybe this somehow helps when also constrained by menu complexity. But it makes \emph{no sense} to instead set price $1/2$: this only decreases the revenue achieved on day $1$, \emph{and} provides stricter constraints on future days (as now she has the option to get the item on day $1$ at a cheaper price).

For our example, we first show that all clean auctions that maintain a good approximation ratio must have high menu complexity. We then follow up by making the claims in the previous paragraph formal: any arbitrary auction of low menu complexity can be derived by ``muddling'' a clean auction, a process which never increases the revenue. A little more specifically, cleaning the menu for deadline $i$ can only increase the revenue and allow more options on later deadlines, without increasing the menu complexity. Formal definitions and claims related to this appear in Appendix~\ref{app:LBA}. We conclude with a formal statement of our lower bound, which proves \autoref{thm:lbmain}.

 \begin{theorem} \label{thm:lowerbound} Any mechanism for the FedEx instance described in \autoref{sec:approxinstance} that has at most $n/8$ menu options on a day $i \in (n/4, n/2]$ has revenue at most $\mathsf{OPT}\left(1-\frac{1}{200000n^2}\right)$.
 \end{theorem}

\section{Conclusions and Future Work}
We provide the first nearly-tight quantitative results on menu complexity in a multi-dimensional setting. Along the way, we design new polygon approximations for a hybrid additive-multiplicative guarantee that turns out to be just right for our application (as evidenced by the nearly-matching lower bounds obtained from the same ideas).

There remains lots of future work in the direction of menu complexity, most notably the push for tighter quantitative bounds in ``truly'' multi-dimensional settings, where the gaps between upper (exponential) and lower (polynomial) are vast. We believe that continuing a polygon approximation approach is likely to yield fruitful results. After all, there is a known connection between concave functions and any mechanism design setting via utility curves, and low menu complexity exactly corresponds to piece-wise linear utility curves with few segments. Still, there are two serious barriers to overcome: first, these utility curves are now multi-dimensional instead of single-dimensional revenue curves. And second, the relationship between utility curves and revenue is somewhat odd (expected revenue is equal to an integral over the support of $\vec{x} \cdot \Delta_f(\vec{x}) - f(\vec{x})$), whereas the relationship between revenue curves and revenue is more direct. There are also intriguing directions for future work along the lines of one-and-a-half dimensional mechanism design, the most pressing of which is understanding multi-bidder instances (as all existing work, including ours, is still limited to the single-bidder setting). 
\section{Instances with regular distributions may require randomness}
\label{sec:reg3daysplit}
For single-dimensional settings, it's well-understood that ``the right'' technical condition on value distributions to guarantee a simple optimal mechanism is \emph{regularity}. This guarantees that ``virtual values'' are non-decreasing and removes the need for ironing, even for multi-bidder settings. Interestingly, ``the right'' technical condition on value distributions to guarantee a simple optimal mechanism for 1.5 dimensional settings is no longer regularity, but \emph{decreasing marginal values}. For example, if all marginals satisfy decreasing marginal values, the optimal mechanism is deterministic for the FedEx problem~\cite{Fiat:2016}, selling a single item to a budget-constrained buyer~\cite{CheG00,DevanurW17}, and a capacity-constrained buyer~\cite{DevanurHP17}.

Still, regularity seems to buy \emph{something} in these problems. For instance, Fiat et al. show that when there are only two possible deadlines, regularity suffices to guarantee that the optimal mechanism is deterministic. It has also been known since early work of Laffont and Robert that regularity suffices to guarantee that the optimal mechanism is deterministic when selling to a budget-constrained buyer with only one possible budget~\cite{LaffontR96}. But the extent to which regularity guarantees simplicity remained open (and was explicitly stated as such in~\cite{DevanurW17}). In this section, we show that regularity guarantees nothing beyond what was already known. In particular, there exists an instance of the FedEx problem with three possible deadlines where all marginals are regular but the optimal mechanism is randomized. This immediately implies an example for a budget-constrained buyer and three possible budgets as well (for instance, just set all three budgets larger than $v_{\max}$ so they will never bind).

We proceed by describing now our instance of the FedEx problem where the optimal auction is randomized, despite all marginals being regular and there only being $3$ possible deadlines (recall that Fiat et al. show that the optimal auction remains deterministic for regular marginals and $2$ deadlines).Throughout this section, instead of using revenue curves $R(v)$, we use $\Gamma(v) = -R(v)$. This is in accordance to~\cite{Fiat:2016}.

\subsection{The setting}
Consider bidders with types distributed as $F_1$ on day $1$, $F_2$ on day $2$, and $F_3$ on day $3$.
\begin{align*}  
F_1(v) &=1 - \mathrm{e}^{-v}, &f_1(v) &= \mathrm{e}^{-v},\\
F_2(v) &=1 - \mathrm{e}^{-5v}, & f_2(v)&=5\mathrm{e}^{-5v},\\
F_3(v) &= 1 - \mathrm{e}^{-\frac{v}{5}}, & f_3(v)&=\frac{1}{5}\mathrm{e}^{-\frac{v}{5}}.\\
\end{align*}

The distribution over days $q$ is  
$$q_1 = \frac{10}{21} \qquad q_2 = \frac{10}{21} \qquad q_3 = \frac{1}{21}.$$

Note that the three distributions are regular but don't have decreasing marginal revenues.

\begin{figure}[htp]
  {\includegraphics[scale = 0.2]{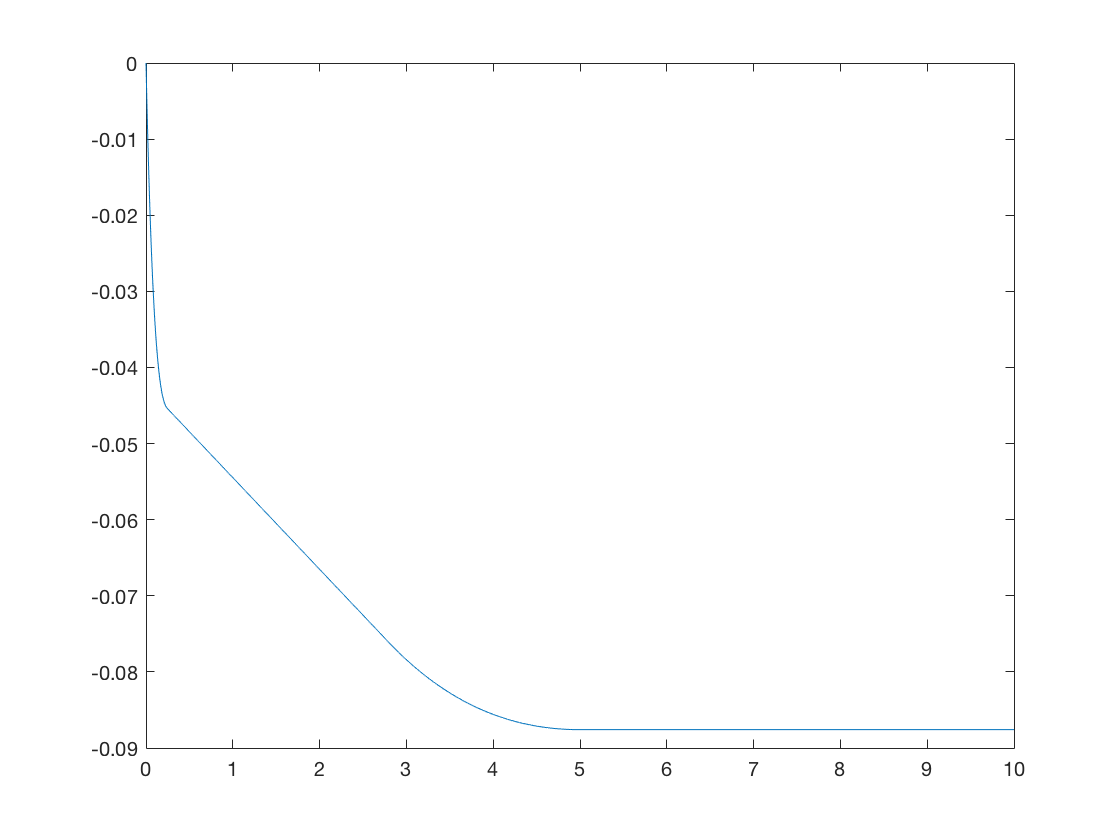}}
  {\includegraphics[scale = 0.2]{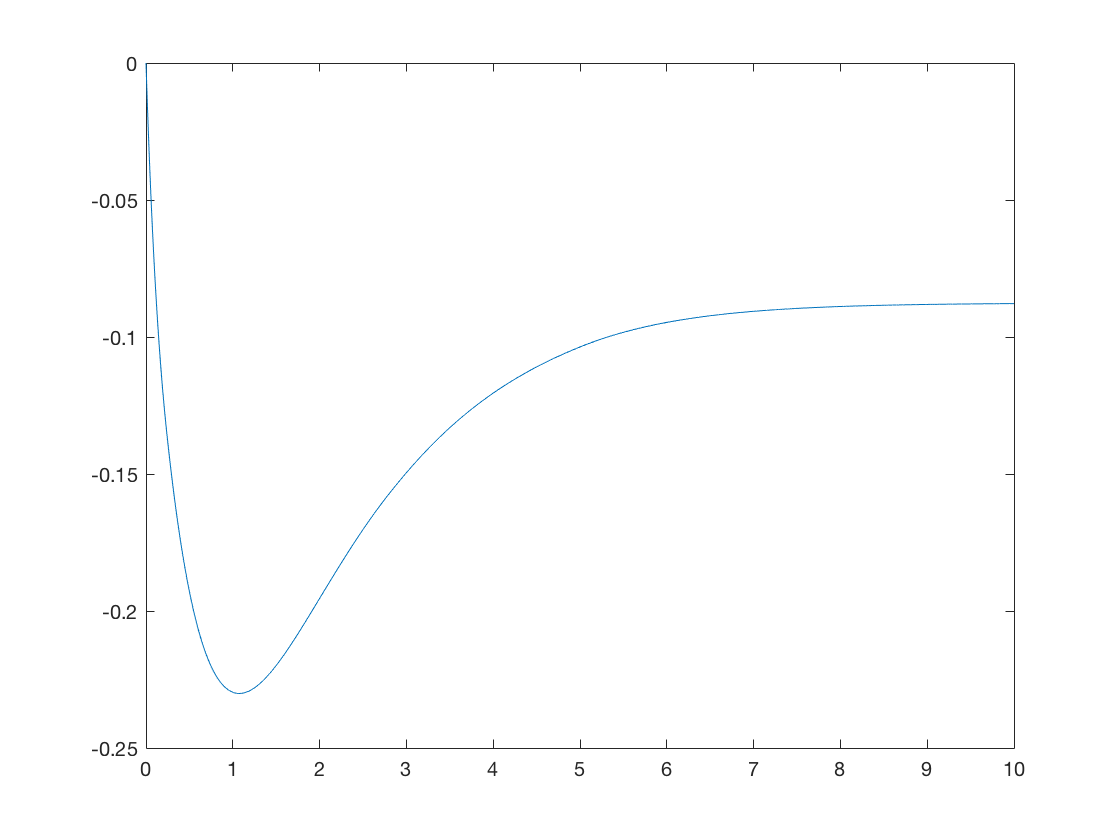}}\\
    \caption{(Left) The curve $\hat{\Gamma}_{\geq 2}$ in our example. (Right) The curve $\Gamma_{\geq 1}$. Note that this curve is minimized at a point in the ironed interval of $\hat{\Gamma}_{\geq 2}$.}
  \label{fig:revcurves}
\end{figure}

\subsection{Analysis}
We use the iterative procedure described in~\cite{Fiat:2016} to find the optimal auction.
\begin{align*}
\Gamma_{\geq 3} = -\frac{v}{21}(1-F_3(v))&= -\frac{v}{21}\mathrm{e}^{-\frac{v}{5}},\\
   r_{\geq 3} &= 5. \\   
\Gamma_{\geq 2} = \begin{cases} 
      \Gamma_2(v) + \hat{\Gamma}_{\geq 3}(v), &  0\leq v \leq 5 \\
      \Gamma_2(v) -\frac{5}{21\mathrm{e}}, & 5 \leq v \\
   \end{cases}&= \begin{cases} 
      -\frac{10v}{21}\mathrm{e}^{-5v}-\frac{v}{21}\mathrm{e}^{-\frac{v}{5}}, &  0\leq v \leq 5 \\
      -\frac{10v}{21}\mathrm{e}^{-5v}-\frac{5}{21\mathrm{e}}, & 5\leq v \\
   \end{cases}  \\
   r_{\geq 2} &\approx 5.0\cdots \\   
\Gamma_{\geq 1} = \begin{cases} 
      \Gamma_1(v) + \hat{\Gamma}_{\geq 2}(v), &  0\leq v\leq 5.0\cdots \\
      \Gamma_1(v) -\frac{5}{21\mathrm{e}}, &  5.0\cdots \leq v \\
   \end{cases}&= \begin{cases} 
            -\frac{10v}{21}\mathrm{e}^{-v}-\frac{10v}{21}\mathrm{e}^{-5v}-\frac{v}{21}\mathrm{e}^{-\frac{v}{5}}, &  0\leq v\leq 0.245\cdots \\
      -\frac{10v}{21}\mathrm{e}^{-v}-\text{\bf A line. See \autoref{fig:revcurves}}, & \bm{0.245\cdots \leq v\leq 2.79\cdots} \\
       -\frac{10v}{21}\mathrm{e}^{-v}-\frac{10v}{21}\mathrm{e}^{-5v}-\frac{v}{21}\mathrm{e}^{-\frac{v}{5}}, & 2.79\cdots\leq v \leq 5.0\cdots \\
      -\frac{10v}{21}\mathrm{e}^{-v}- 0.0876\cdots, & 5.0\cdots\leq v \\
   \end{cases}  \\
   r_{\geq 1} &\approx 1.07. \\   
 \end{align*}
All real values written are approximate and end with ``$\cdots$''. The case in {\bf boldface} denotes the ironed interval. The optimal price for day $1$ is $1.07$ which is in that interval. Hence the optimal auction has to be randomized.

%Bibliography

\begin{appendices}

\section{Additional preliminaries}
\label{app:preliminaries}

In this appendix, we summarize the approach in \cite{Fiat:2016} to obtain the optimal mechanism for an arbitrary ``FedEx'' instance. We begin with the linear program that encodes this optimization problem.
\begin{equation}
\label{eq:LP1}
\begin{aligned}
& {\text{maximize}} 
& \sum_{i} \sum_{v} q_i f_i(v) p(v,i)  \\
& \text{subject to}
&  \pi(v,i) v - p(v,i) & \geq \pi(v-1, i) v - p(v-1, i) & \forall v \geq 1, i  & & \textrm{(Leftwards IC)} \\ 
& & \pi(v,i) v - p(v,i) &\geq \pi(v+1, i) v - p(v+1, i) & \forall v < v_{max}, i  & & \textrm{(Rightwards IC)} \\ 
& & \pi(v,i) v - p(v,i) & \geq \pi(v, i-1) v - p(v, i-1) & \forall i > 1, v  & & \textrm{(Downwards IC)} \\ 
& & \pi(v,i) & \leq 1 & \forall i, v  & & \textrm{(Feasibility)} \\ 
& & \pi(0,1) & = p(0,1) = 0. & & & \textrm{(Individual Rationality)}
\end{aligned}
\end{equation}

Note that we have not included the constraints where the bidder misreports a higher deadline. No rational bidder would consider these deviations since they would always get non-positive utility. We now formally present the allocation curves described in Section~\ref{sec:definitions}. This, combined with the definitions of optimal revenue curves, provide a clean characterization of optimal auctions for any instance of the FedEx problem.

\begin{definition}[Optimal allocation curves~\cite{Fiat:2016}]  \label{def:optcurves}

Let $j^*$ be the largest $j$ such that $v_j \leq r_{\geq i}$. For any $j \leq j^*$, consider two cases: 

\begin{itemize}
	\item $\hat{R}_{\geq i}$ is not ironed at $v_j$. Then 
		\[  a_{i,j}(v) =  \left\{ 
\begin{array}{ll}
      0 & v < v_j \\
      1 & \text{else}. \\
\end{array} 
\right. \]
	\item $\hat{R}_{\geq i}$ is ironed at $v_j$. Let $\underline{v}_j$ be the largest $v < v_j$ such that $R_{\geq i}$ is not ironed at $v$ and $\overline{v}_j$ be the smallest $v_j < v$ such that $R_{\geq i}$ is not ironed at $v$. Let $0 < \delta < 1$ be such that 
	$$ v_j = \delta \underline{v}_j + (1-\delta)\overline{v}_j. $$
	Define 
	\[  a_{i,j}(v) =  \left\{ 
\begin{array}{ll}
      0 & v < \underline{v_j} \\
      \delta & \underline{v_j} \leq v \leq \overline{v_j} \\ 
      1 & v > \overline{v_j}. \\
\end{array} 
\right. \]
\end{itemize}

Then set $a_i(v)$ as follows 

	\[  a_{i}(v) =  \left\{ 
\begin{array}{ll}
      \sum_{j=1}^{j*} (\beta_j - \beta_{j-1}) a_{i,j}(v) & v < r_{\geq i} \\
      1 & v \geq r_{\geq i}, \\
\end{array} 
\right. \]

where $\beta_j$ is the probability of allocating the item with value $a_j$ on day $i-1$. 

\end{definition}

\begin{lemma}~\cite{Fiat:2016}
\label{lem:optimalauction}
The allocation curves $a_i$ are monotone increasing from $0$ to $1$ and satisfy the Downwards IC from~\ref{eq:LP1}. Moreover, each $a_i$ changes value only at points where $\hat{R}_{\geq i}$ is not ironed. 
\end{lemma}

\begin{remark}
Every solution to LP~\ref{eq:LP1} is an optimal mechanism for the FedEx problem.  
\end{remark}

In addition, we present this simple claim, which is unrelated to the FedEx problem itself, that will be useful in future sections. 

\begin{claim}
\label{lem:virtualrev}
Let $f$ be a distribution over $[v_{\max}]$, and let $\phi_f, R_f$ be the corresponding virtual value and revenue curve, respectively. Then, for all $v \in [v_{max}-1]$ 
$$R_f(v) - R_f(v-1) = \phi_f(v) f(v) .$$ 
\end{claim}

\begin{proof}
This follows from some simple calculations: 

\begin{equation*}
\begin{aligned}
R_f(v) - R_f (v-1) & = v(1-F(v-1))-(v+1)(1-F(v)) \\
& = v(F(v)-F(v-1)) - (1-F(v)) \\
& = vf(v) - (1-F(v)) =  f(v) \phi_f(v) .
\end{aligned}
\end{equation*}
\end{proof}

\section{Omitted proofs and perturbed example of Section~\ref{sec:LB}}
\label{sec:optworstcase}

\subsection{Omitted proofs}

\begin{lemma}
\label{lem:makingrev}
Fix $v > 0$. Let $(2v)^{-1} > \varepsilon > 0$ and let $R = \{r_i\}_{i=1}^{v}$ be a sequence of real numbers such that $\lvert{r_{i+1} - r_i}\rvert \leq \varepsilon, \forall i \in [v-1]$. If $r_1 = 1	$, there exists a distribution $f: [v] \to [0,1]$  such that $R_f(i) = r_i, \forall i \in [v]$.   
\end{lemma}

\begin{proof}
By our choice of $\varepsilon$, all elements in the sequence are greater than $1/2$. Let $r_{v+1} = 0$ and define

$$F(i) = 1 - \frac{r_{i+1}}{i+1}, \forall i \in \{0,1,2,\cdots, v\}.$$ 

We will show that $F$ indeed corresponds to a valid distribution function by showing that it is non-negative and non-decreasing. Once we have shown this, it becomes clear that $R_f(i) = r_i$, proving the Lemma.

\begin{claim}
\label{claim:nonneg}
$F(i)$ is non-negative, non-decreasing. 
\end{claim} 

\begin{proof}
For $i = 0$, $F(i) = 1-1 = 0$. Suppose $i \geq 1$. Since $0 \leq r_i < 2, \forall i$, it follows that $1\geq F(i) \geq 1-\frac{r_{i+1}}{2} > 0$. To show it is non-decreasing, consider the difference between two consecutive terms. For all $v > i \geq 1$,

\begin{equation*}
\begin{aligned} 
F(i) - F(i-1) & = \frac{r_i}{i} - \frac{r_{i+1}}{i+1} = \frac{r_i}{i(i+1)} + \frac{r_i - r_{i+1}}{i+1} \\ 
& \geq \frac{1}{i+1} \left(\frac{r_i}{i} - \varepsilon \right) \\ 
& > \frac{1}{i+1} \left(\frac{r_i}{v} - \frac{1}{2v} \right) \geq 0. 
\end{aligned} 
\end{equation*}

For $i = v$, $F(i) - F(i-1)  = \frac{r_i}{i}  \geq 0. $

\end{proof}
\end{proof}

We are now ready to explicitly construct an example that achieves deadline menu complexity of $2^{n-1}$ where $v_{\max} = 2^{n}-1$. Fix $n$ and construct $n-1$ sequences $S_i = \{s_{ij}\}_{j=1}^{2^{n}-1}$ of length $2^{n}-1$, then 

$$s_{ij} = \varepsilon \cdot (1-2b_{i+1}(j)) b_i(j) \left( \prod_{k=1}^{i-1} (1-b_k(j)) \right),$$

where $b_i(j)$ denotes the $i$-th least significant bit in the binary expansion of $j$, $\varepsilon = 4^{-n}$. Let $S_n = \{s_{nj} \}_{j=1}^{2^{n}-1}$ as $s_{nj} = \varepsilon (2b_n(j)-1)$. Lemma~\ref{lem:makingrev} implies that there exist distributions $f_j$ $\forall j \in [n]$ such that $\phi_{f_j} f(j) = s_i \forall i \in [v-1]$. In our construction the type distribution of day $i$ corresponds to the distribution $f_{n+1-i}$. Let the distribution over days be uniform (i.e. $q_i = \frac{1}{n} \forall i \in [n]$).	 In order to show that this construction achieves a menu complexity of $2^{n-1}$ we first need to characterize the revenue curves for all days, and then show that an optimal auction exists where prices can be set so as to create a large menu. The intuition for these revenue curves is that their ironed intervals are nested: prices at the endpoints of ironed intervals on day $i$ are the midpoints of new ironed intervals on day $i+1$. In addition, after ironing the curves, they look like constants. Therefore, the optimal revenue curves will have the same ironed intervals as their original counterparts. We design the revenue of the first day to be maximized at the median value. On the next day this price will belong to an ironed interval, meaning that any optimal auction must offer a lottery over two prices that are not ironed for day 2. The size of the lottery offered directly translate into the minimum number of options a menu for that day must have. By the nesting construction, this will double the number of options offered on day $i+1$ with respect to day $i$. 

We can now use the above construction, combined with Claim~\ref{lem:virtualrev} and Lemma~\ref{lem:makingrev}, to show a more general result, Lemma~\ref{optrevexp}. 

\begin{proof}[Proof of \autoref{optrevexp}]

By simple examination, the revenue curve for day $1$ is just a line that increases until it reaches $v_{\max}/2$ and then decreases, so it is maxed at $v_{\max}/2$. Consider day $i \geq 2$. The sequence that generates its revenue curve, $s_{n+1-i,j}$ is non-zero iff $j$ has remainder $2^{n-i} \mod 2^{n-i+1}$. Since there are $2^{n}-1$ values in the sequence, there will be $2^{i-1}$ non-zero values. These values will alternate between $\varepsilon, -\varepsilon$ and each alternation creates one ironed interval: the revenue decreases by $\varepsilon$, stays at that value for a while, and increases by $\varepsilon$ again. This gives $2^{i-2}$ ironed intervals for the revenue curve of day $i$. Moreover, the last price where the revenue increases is at $2^{i}(2^{n-i+1}-1)$, making it a valid maximizer. The revenue remains constant from there on so any higher value is also a maximizer. 

For the third point, note that $R_{i}$ takes values in $\{1, 1-\varepsilon \}$, alternating between them for intervals whose lengths depend on $i$. $\hat{R}_i$, the upper concave hull of $R_i$, is then a constant function with value $1$ everywhere.  

We prove the last point by induction, starting from $R_{n}$. This holds because $R_{n} = R_{\geq n}$. Note that, from our previous claim, $\hat{R}_{\geq n} = \hat{R}_{n}$ is a constant function. Suppose that $R_{\geq i+1}(v) = R_{i+1}(v) + c$ for some $c$. For $v \leq r_{\geq i+1}$, $R_{\geq i}(v) = R_{i}(v) + \hat{R}_{\geq i+1}(v)$. For $v > r_{\geq i+1}$, $R_{\geq i}(v) = R_{i}(v) + \hat{R}_{\geq i+1}(r_{\geq i+1})$. In either case, the term added to $R_i(v)$ is a constant (and it is the same constant) by the inductive hypothesis, so the claim follows. 
\end{proof}

\begin{lemma} 
\label{optallocexp}
The series of optimal revenue curves $R_{\geq i}$ induced by the distributions $f_{n+1-i}$ is such that on any day $i$, \emph{an} optimal allocation curve as constructed by~\cite{Fiat:2016} is a step function with $2^{i-1}$ jumps and takes the following form: 
	\[  a_{i}(v) =  \left\{ 
\begin{array}{ll}
      0 & v < 2^{n-i}, \\
      \frac{k+1}{2^{i-1}} & 2^{n-i}+k2^{n-i+1} \leq v < 2^{n-i}+(k+1)2^{n-i+1}, \\ 
      1 & v \geq 2^{n}-2^{n-i}, \\
\end{array} 
\right. \]	
	where $0 \leq k \leq 2^{i-1}-2$.  
	%jumps. Namely it will be change values at prices $\{2^{d-i} + k 2^{d-i+1}\}_{k=0}^{2^{i-1}-1}$. 
	Moreover, these prices where the function jumps on day $i$ will belong to ironed intervals for the optimal revenue curve of the following day, $R_{\geq i+1}$ for all $i < n$. 
\end{lemma}

\begin{proof}
We prove this also by induction going from the first day to the last. As stated before, $R_{\geq 1}$ is maximized at $v_{\max}/2$ and has no ironed intervals, therefore it is clear $r_{\geq 1} = v_{\max}/2$ and the optimal allocation curve $a_1$ corresponds to a step function at $v_{\max}/2$. By construction, this price belongs to an ironed interval of $R_{\geq 2}$. Thus the base case holds. Suppose that the statement is true for day $i$. We want to understand the optimal allocation curve for day $i+1$.

First note that the places where the function jumps on day $i$, $P_i = \{2^{n-i}+k2^{n-i+1}\}_{k=0}^{2^{i-1}-1}$, belong to ironed intervals of $R_{\geq i+1}$ (which are the same intervals as  $R_{i+1}$). This is because at these prices $j \in P_i$ the function $s_{n-i+1, j}$ takes a value of $0$. The nearest place where they are non-zero are exactly at $j-2^{n-i-1}$ and $j+2^{n-i-1}$, where the function takes values $\varepsilon$ and $-\varepsilon$, meaning we are inside an iron interval (the revenue decreases and then increase by $\varepsilon$). Thus by the optimal allocation curves suggested in~\cite{Fiat:2016} we observe that if on day $i$ we offered a price of $2^{n-i}+k 2^{n-i+1}$ with positive probability $\frac{k+1}{2^{i-1}}$, we must also allocate at prices $2^{n-i-1}+(2k)2^{n-i}$ and $2^{n-i-1}+(2k+1)2^{n-i}$ on day $i+1$ with positive probability. 

The probability for allocating the item with price $2^{n-i-1}+ k 2^{n-i} \leq v < 2^{n-i-1}+(k+1) 2^{n-i}$ depends on the parity of $k$. If $k$ is odd then the values in this range correspond to a non-ironed interval on day $i+1$, meaning they preserve the probability of allocation from day $i$. The probability of allocating on the interval with endpoints $2^{n-i}+\frac{(k-1)}2^{n-i+1}$ and $2^{n-i}+\frac{k+1}{2}2^{n-i+1}$, which contains our new interval of interest, is $\frac{k+1}{2^{i+1}}$. If $k$ is even then we belong to an ironed interval on day $i+1$, meaning that the probability of allocation is going to be the average of allocating on the two intervals on day $i$ that intersect this one. These intervals are $[2^{n-i}+\frac{k-2}{2} 2^{n-i+1}, 2^{n-i}+\frac{k}{2} 2^{n-i+1}]$ and $[2^{n-i}+ \frac{k}{2} 2^{n-i+1}, 2^{n-i}+\frac{(k+2)}{2} 2^{n-i+1}]$. Thus the probability of allocating at $v$ on day $i+1$ is just $\frac{k+1}{2^{i+1}}$.
\end{proof}

\begin{figure}[htp]
	\centering
	\includegraphics[scale = .75]{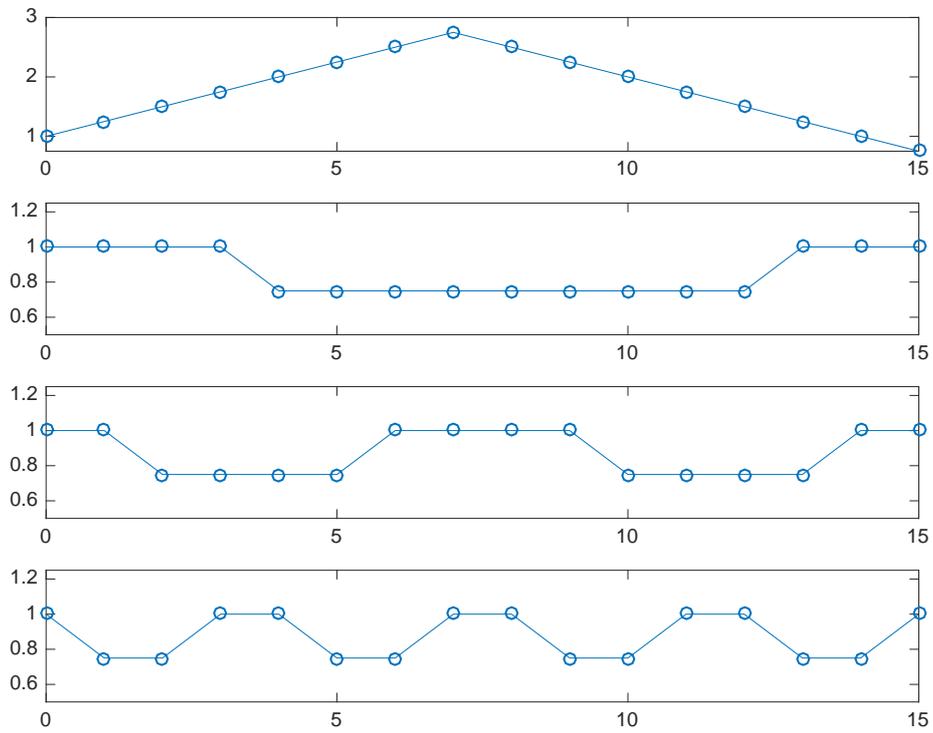}
	\caption{Revenue curves corresponding to an instantiation of our construction with $n = 4$ and $\varepsilon = .25$.}
	\label{fig1:fourdayrev}
\end{figure}

With this understanding of the revenue curves of the instance we construct, we are ready to prove Theorem~\ref{thm:expmenu}, the main result of Section~\ref{sec:LB}. 

\begin{proof}[Proof of Theorem~\ref{thm:expmenu}]
By Lemma~\ref{optallocexp}, the optimal allocation curves for day $i$ are step-wise increasing functions with $2^{i-1}$ steps. This means that for any deadline $i$, the $i$-deadline menu complexity of this instance is $2^{i-1}$. Combining the number of options offered over all $n$ days gives a menu complexity of $2^{n}-1$.  
\end{proof}

\subsection{Perturbed case}

In this appendix, we show how to tweak the example in Section~\ref{sec:LB} so that {\em no} optimal auction has a menu complexity less than $2^n$.  The problem with the example in Section~\ref{sec:LB} is that we don't have to follow the allocation suggestion of~\cite{Fiat:2016} in order to achieve an optimal auction because we could simply choose larger ironed intervals on every day that spanned the whole spectrum of prices and (because of the simplicity of the construction) still recover all the revenue. We add a small non-linear term to the revenue curve of each day to dissuade from this while still preserving the 'nested' structure of ironed intervals. Consider now the $n-1$ sequences 

$$s_{ij} = \frac{\varepsilon}{2} \cdot (1-2b_{i+1}(j)) b_i(j) \left( \prod_{k=1}^{i-1} (1-b_k(j)) \right) + \delta j \cdot (2 v_{\max} - j) ,$$

where $v_{\max} = 2^n, \epsilon = \frac{1}{4v}$ and $\delta$, the weight of the non-linear term, is $v^{-10}$. The distribution for the first day is the same as the one we used for the previous case. Note that the non-linear term added is maxed at $v_{\max}$. Again, Lemma~\ref{lem:makingrev}, Claim~\ref{lem:virtualrev} allow us to conclude there are revenue curves $R_i$ and valid distributions $f_{n+1-i}$, whose changes are dictated by the sequence $s_{n+1-i}$. We restate the results from~\ref{sec:LB} with the appropriate adjustments. 

\begin{lemma}
\label{optrevexp2}
The series of revenue curves $R_{i}$ induced by the distributions $f_{n+1-i}$ satisfy the following properties: 
\begin{itemize}
	\item $R_1$ is maximized at $v_{\max}/2$ and has no ironed intervals. 
	\item $R_i$ for $1 < i \leq d$ has a unique maximizer at price $v_{\max}$ and has $2^{i-2}$ ironed intervals. 
	\item $R_{\geq i}$ has the same ironed intervals as $R_{i}$. % In fact, $R_{\geq i}$ is just $R_{i}$ except at $v_{\max}$.  
	\end{itemize}
\end{lemma} 

\begin{proof}
The first point remains true since we haven't changed $R_1$. Sine $v_{\max}$ was a maximizer of the function before adding the non-linear term, it will remain a maximizer since it's also optimal for that function. In fact, it is now the \emph{unique} maximizer. This implies that $r_{\geq n} = v_{\max}$. The last point follows from a similar inductive proof to the one in the original case. It is easy to see that now $R_{\geq i}$ will be maximized at $v_{\max}$, for all $i$. Then by the characterization from~\cite{Fiat:2016} we have that for any $v < v_{\max}$, $R_{\geq i}(v) = R_{i}(v)$. 
\end{proof}

The key property of Lemma~\ref{optallocexp} is that the ironed intervals for day $i$ are nested in those of day $i+1$. This property is kept despite the small adjustment of the function, so the statement of the Lemma remains and the proof of  the main theorem of Section~\ref{sec:LB} follows. After the adjustment the ironed intervals become unique and no clever algorithm can take large ironed intervals to avoid a large menu complexity.
\section{Omitted proofs of Section~\ref{sec:UB}}
\label{app:UB}

\subsection{Polygon approximation of concave functions} \label{sec:polygonapproximation}

Polygon approximation of functions is a classic problem in approximation theory. The central problem in approximation theory is to see how `well' does a class of `simple' functions approximate an arbitrary function. Usually, the `simple' functions have a property that makes them easy to study and the approximation scheme ensures that results carry over to arbitrary functions.  In the subfield of polygon approximation, the class of `simple' functions is the set of all continuous piecewise linear functions. Different error metrics (an error metric defines the properties of a good approximation) are studied, the most common being an additive error defined by a parameter $\varepsilon$. We define polygon approximation formally and state a celebrated result for additive $\varepsilon$ error. 

\begin{definition}[Polygon approximation] \label{def:polygonapproximation} Given a function $f:[a,b] \to \mathbb{R}$ and a sequence $X$ of points $a= x_0, x_1, \cdots, x_{n-1}, x_n = b$,  define $\tilde{f}_X$, the $X$-approximation of $f$, to be the continuous piecewise linear function that passes through $(x_0,f(x_0)), (x_1,f(x_1)), \cdots, (x_n, f(x_n))$.
\end{definition}

Before continuing, we briefly state the notation that we are going to use throughout this section. For a concave function $f$, we use $f^+$ to denote the right-derivative and $f^-$ to denote the left derivative. We remind the reader that for a concave function over an interval, both $f^-$ and $f^+$ are well defined at every point (and there are only countably many points for which $f$ is not differentiable). The following theorem has been taken from \cite{Rote91} and has been reworded to fit our notation. %We make some assumptions ({\em e.g.} that $f^+(0) \leq 1$ and $f^-(1)\geq 0$) that are without loss of generality in our setting of the FedEx problem, but the reader familiar with polygon approximation may be used to seeing these statements in a more general form. 

\begin{theorem}[\cite{Rote91,Burkard91}] \label{thm:rote91} For any $\varepsilon >0$ and concave $f:[a,b] \to \mathbb{R}$ such that $b-a = L$ and $f^+(a) -f^-(b) \leq \Delta$, there exists a sequence  $X$ of at most $3+\sqrt{\frac{9}{8}\frac{L\Delta}{ \varepsilon}}$ points such that $\forall x \in [a,b], 0\leq f(x) - \tilde{f}_X(x)\leq \varepsilon$.
\end{theorem}

\autoref{thm:rote91}, as stated above, deals with arbitrary concave functions  and is known to be tight~\cite{Rote91}. In our setting of the FedEx problem, the concave functions that we wish to approximate are monotone revenue curves defined over $[0,r_{\geq i}]$ (for the rest of this section, we will use $v_{\max}$ instead of $r_{\geq i}$ to refer to the right end-point of the interval to keep the notation non-specific to the FedEx problem). Being monotone revenue curves, they satisfy many other properties, {e.g.}, $f(0) = 0$, $f^+(x) \leq 1$, and $f^-(x)\geq 0$. Also, the error that we are interested in is {\em not} the additive $\varepsilon$ error that is dealt with in the theorem. \autoref{lemma:epsilonpolygonapproximation} formalizes  the exact guarantees we desire. Note that the error metric that we use is an additive multiplicative hybrid that increases with $f(v_{\max})$. To prove this lemma, we use \autoref{thm:rote91} separately on each of the intervals $[0,1],[1,2], \cdots,[v_{\max}/4,v_{\max}/2],[v_{\max}/2,v_{\max}]$. We show that when the size of the approximating sequence $X$ is large in these intervals, then $f(v_{\max})$ must be large. In turn, this implies that we have more wiggle room in our error. We exploit this room to improve the dependence on $v_{\max}$ to logarithmic from the polynomial dependence on $L$ in \autoref{thm:rote91}.

\subsection{Proof of Theorem~\ref{thm:polygonapproximation}}
We begin with an easy lemma: \autoref{lemma:fvlowerbound} shows how to lower bound $f(v_{\max})$ using the derivatives of various points along the way. 

\begin{lemma}\label{lemma:fvlowerbound} For any concave $f:[0,v_{\max}] \to [0,\infty)$ such that $f^-(v_{\max}) \geq 0$, we have 
\[f(v_{\max}) \geq \sum_{i=0}^{\lceil\log v_{\max}\rceil}f^+\left(\frac{v_{\max}}{2^{i}}\right)\frac{v_{\max}}{2^{i+1}}.\]
\end{lemma}
\begin{proof}
Since the function $f$ is concave, we get that for all $i > 0 $,  

\[f\left(\frac{v_{\max}}{2^{i-1}}\right) - f\left(\frac{v_{\max}}{2^{i}}\right) \geq f^+\left(\frac{v_{\max}}{2^{i-1}}\right)\frac{v_{\max}}{2^i}.\]

%Also, we have 
%\[f\left(\frac{v_{\max}}{2^{\lceil{\log v_{\max}}\rceil}}\right)  \geq f^+\left(\frac{v_{\max}}{2^{\lceil{\log v_{\max}}\rceil}}\right)\frac{v_{\max}}{2^{\lceil{\log v_{\max}}\rceil}}.\]

Adding these inequalities for $i = 1,2,\cdots \lceil{\log v_{\max}}\rceil + 1$, we get 
\[f(v_{\max})\geq \sum_{i=1}^{\lceil\log v_{\max}\rceil + 1}f^+\left(\frac{v_{\max}}{2^{i-1}}\right)\frac{v_{\max}}{2^i} \geq \sum_{i=0}^{\lceil\log v_{\max}\rceil}f^+\left(\frac{v_{\max}}{2^{i}}\right)\frac{v_{\max}}{2^{i+1}}.\]
\end{proof}

%We'll make use of \autoref{thm:rote91} to prove the following improved bound when we're willing to let the error be either absolute or scale with $f(v_{\max})$. 

And now, we show how to improve the depends on $v_{\max}$ from linear to logarithmic for our relaxed hybrid guarantee.

\begin{lemma} \label{lemma:epsilonpolygonapproximation} For any $\varepsilon >0$ and concave function $f:[0,v_{\max}] \to [0,\infty)$ such that $f(0) = 0$, $f^+(0) \leq 1$, $f^-(v_{\max}) \geq 0$, there exists a sequence $X$ of at most $O\left(\log v_{\max} + \sqrt{\frac{\log v_{\max}}{\varepsilon}}\right)$ points such that for all $x \in [0,v_{\max}],$
$$f(x)-\varepsilon \left(1+f(v_{\max})\right) \leq \tilde{f}_X(x) \leq f(x).$$
\end{lemma}

\begin{proof}
We divide the domain of the function $f$ into the intervals $I_i = \Big[\frac{v_{\max}}{2^i},\frac{v_{\max}}{2^{i-1}}\Big]$ for $i\in \{1,2,\cdots\lceil\log v_{\max}\rceil\}$ and the interval $I_0 = \Big[0,\frac{v_{\max}}{2^{\lceil\log v_{\max}\rceil}}\Big]$. For each interval $I_i$, we define a sequence $X_i$ that approximates $f$ well over this interval. The final sequence $X$ is obtained by combining all the sequences $X_i$.

We start with the interval $I_0$. The length of $I_0$ is $\frac{v_{\max}}{2^{\lceil\log v_{\max}\rceil}} \leq 1$. Since $f^+(0) \leq 1$,  we can apply \autoref{thm:rote91} with $\Delta = 1$ and conclude that to approximate $f$ up to an additive $\varepsilon \leq \varepsilon \left(1+f(v_{\max})\right)$ error on this interval, we need a sequence $X_0$ of at most $3+\sqrt{\frac{9}{8 \varepsilon}}$ points.

We now argue for the rest of the domain of $f$. The length $L_i$ of the interval $I_i$ is $\frac{v_{\max}}{2^i}$. Let $\Delta_i = f^+\left(\frac{v_{\max}}{2^i}\right) - f^-\left(\frac{v_{\max}}{2^{i-1}}\right)$. Applying \autoref{thm:rote91} on $I_i$ with the error parameter $\varepsilon f(v_{\max}) \leq \varepsilon \left(1+f(v_{\max})\right) $ gives a sequence $X_i$ at most $3+\sqrt{\frac{9}{8}\frac{v_{\max}\Delta_i}{ 2^i \varepsilon f(v_{\max})}}$ points such that $\tilde{f}_{X_i}$ approximates $f$ on $I_i$ up to an additive~$\varepsilon f(v_{\max})$ error.

Combine all the sequences $X_i$ for $i = 0,1,\cdots, \lceil\log v_{\max}\rceil$ to get a single sequence $X$ such that $\tilde{f}_X$ approximates $f$ up to an additive $\varepsilon(1+f(v_{\max}))$ throughout its domain $[0,v_{\max}]$. The number of points in $X$ is at most 

\begin{align*}
3+\sqrt{\frac{9}{8 \varepsilon}} + &\sum_{i=1}^{\lceil\log v_{\max}\rceil} 3+\sqrt{\frac{9}{8}\frac{v_{\max}\Delta_i}{ 2^i \varepsilon f(v_{\max})}} \\
&\leq  3 (1 + \lceil\log v_{\max}\rceil)+\sqrt{\frac{9}{8 \varepsilon}}+ \frac{1}{\sqrt{\varepsilon f(v_{\max})}}\sum_{i=1}^{\lceil\log v_{\max}\rceil}\sqrt{\frac{9}{8}\frac{v_{\max}\Delta_i}{ 2^i }}\\
& \leq 3 (1 + \lceil\log v_{\max}\rceil)+\sqrt{\frac{9}{8 \varepsilon}}+ \frac{1}{\sqrt{\varepsilon f(v_{\max})}}\sqrt{\lceil \log v_{\max}\rceil\sum_{i=1}^{\lceil\log v_{\max}\rceil}\frac{9}{8}\frac{v_{\max}\Delta_i}{ 2^i }} & \textrm{(Cauchy-Schwarz)}\\
& \leq 3 (1 + \lceil\log v_{\max}\rceil)+\sqrt{\frac{9}{8 \varepsilon}}+ \frac{1}{\sqrt{\varepsilon f(v_{\max})}}\sqrt{\lceil \log v_{\max}\rceil\sum_{i=1}^{\lceil\log v_{\max}\rceil}\frac{9}{8}\frac{v_{\max}f^+\left(\frac{v_{\max}}{2^i}\right)}{ 2^i }} & \textrm{(Definition of $\Delta_i$)}\\
& \leq 3 (1 + \lceil\log v_{\max}\rceil)+\sqrt{\frac{9}{8 \varepsilon}}+ \sqrt{\frac{18}{8} \frac{\lceil\log v_{\max}\rceil}{\varepsilon}} &\textrm{(\autoref{lemma:fvlowerbound})}\\
& = O\left(\log v_{\max} + \sqrt{\frac{\log v_{\max}}{\varepsilon}}\right).\\
%& <  \lceil\log v_{\max}\rceil+ \frac{1}{\sqrt{\varepsilon f(v_{\max})}}\sqrt{\lceil \log v_{\max}\rceil 2f(v_{\max})}  = \lceil\log v_{\max}\rceil+ \sqrt{\frac{2\lceil\log v_{\max}\rceil}{\varepsilon}}. & \textrm{(\autoref{eq:revlowerbound})}\\
\end{align*}

\end{proof}

It is noteworthy that \autoref{lemma:epsilonpolygonapproximation} preserves the quadratic dependence on $\varepsilon$ in \autoref{thm:rote91}, and simply replaces $v_{\max}$ with $\ln(v_{\max})$. Note that essentially the main idea is that we wish to target additive error $\varepsilon \cdot f(v_{\max})$ in the domain of ``large $x$'' (as targeting an additive $\varepsilon$ in this range requires linear dependence on $v_{\max}$ that can be avoided if $f(v_{\max})$ is big) but additive error $\varepsilon$ in the domain of ``small $x$'' (as targeting an additive $\varepsilon $ requires linear dependence on $v_{\max}$ which is small). So in order to get the improved guarantee, our bound needs the flexibility to take either the additive or multiplicative approximation in the different ranges. 

To complete the proof of~\autoref{thm:polygonapproximation}, we additionally provide a much simpler argument showing that the dependence on $v_{\max}$ can be removed at the cost of suboptimal dependence on $\varepsilon$.

%Note also that an alternative (and much simpler) approach yields a polygon approximation of size $1/\varepsilon$, and also doesn't require as many assumptions on $f$.

\begin{lemma}\label{lem:easy}
For any $\varepsilon > 0$ and any increasing concave function $f:[0,v_{\max}] \to [0,\infty)$ such that $f(0) = 0$, there exists a sequence $X$ of at most $O(1/\varepsilon)$ points such that for all $x \in [0,v_{\max}]$, \[f(x)-\varepsilon  f(v_{\max}) \leq \tilde{f}_X(x) \leq f(x).\]
\end{lemma}
\begin{proof}
We divide the range $[0, f_{\max}]$ of the function $f$ into $1/\varepsilon$ intervals and take one point from each interval. Formally, let $x_i = \min_x \{x \mid f(x) \geq  i \varepsilon f(v_{\max}) \}$, for all $i \in \{0,\cdots,\lfloor{1/\varepsilon}\rfloor\}$. Note that since $f$ is continuous (it is concave), $f(x_i) = i \varepsilon f(v_{\max})$. 

We define the sequence $X$ to be the points $\{x_i\}_{i=0}^{\lfloor{1/\varepsilon}\rfloor}$ followed by $x_{\lfloor{1/\varepsilon}\rfloor + 1} = v_{\max}$. We prove that $\tilde{f}_X$ is the required polygon approximation. For any point $x \in [0,v_{\max}]$, let $i \in \{0,1,\cdots, \lfloor{1/\varepsilon}\rfloor\}$ be such that $x_i \leq x < x_{i+1}$. Since $f$ is increasing, we have $f(x_i) \leq f(x) \leq f(x_{i+1})$. This gives $ i \varepsilon f(v_{\max}) \leq f(x) \leq  (i+1) \varepsilon f(v_{\max})$.

Since $\tilde{f}_X(x)$ is a piecewise linear function, we have $f(x_i)  = \tilde{f}_X(x_i)\leq \tilde{f}_X(x) \leq \tilde{f}_X(x_{i+1}) = f(x_{i+1})$. Again, we get $ i \varepsilon f(v_{\max}) \leq \tilde{f}_X(x) \leq  (i+1) \varepsilon f(v_{\max})$.

Combining the two inequalities, we get  $f(x) - \tilde{f}_X(x) \leq \varepsilon f(v_{\max})$. Since $f$ is concave, any piecewise linear segment joining two points on the function lies below the function and we have:

\[0 \leq  f(x) - \tilde{f}_X(x) \leq \varepsilon f(v_{\max}).\] 

\end{proof}

Given any function $f$, we can use either of the two polygon approximation schemes defined by \autoref{lemma:epsilonpolygonapproximation} and \autoref{lem:easy}. Taking the better of these two schemes we get a proof of~\autoref{thm:polygonapproximation}. 
%Combining both lemmas together, we obtain the following theorem (whose proof essentially observes that we can take a minimum more cleverly than just $\min\{1/\varepsilon, \log v_{\max} + O(\sqrt{\frac{\log v_{\max}}{\varepsilon}})\}$). 
\begin{proof}[Proof of \autoref{thm:polygonapproximation}]
Let $X_1$ and $X_2$ be the sequences given by \autoref{lemma:epsilonpolygonapproximation} and \autoref{lem:easy} respectively. Suppose $1/\varepsilon \leq \log v_{\max}$. In this case, the size of $X_1$ is at most $O(1/\varepsilon) = O\left(\min\left\{1/\varepsilon,\sqrt{\frac{\log v_{\max}}{\varepsilon}}\right\}\right)$. Now suppose $1/\varepsilon \geq \log v_{\max}$. In this case, the size of $X_2$ is at most $O\left(\log v_{\max} + \sqrt{\frac{\log v_{\max}}{\varepsilon}}\right) = O\left(\sqrt{\frac{\log v_{\max}}{\varepsilon}}\right) = O\left(\min\left\{1/\varepsilon,\sqrt{\frac{\log v_{\max}}{\varepsilon}}\right\}\right)$. Thus, the smaller of the two sequence is of size at most $O\left(\min\left\{1/\varepsilon,\sqrt{\frac{\log v_{\max}}{\varepsilon}}\right\}\right)$.
\end{proof}

%In the next subsection, we prove that this bound is tight in the following sense: first, if you want to remove dependence on $v_{\max}$ entirely, then $1/\varepsilon$ is the best you can hope for. Second, if you are willing to depend both on $v_{\max}$ and $\varepsilon$, then the bound in Theorem~\ref{thm:polygonapproximation} is tight.

\subsection{Proof of Proposition~\ref{prop:worstexample}} 
\label{proof:tight}

\begin{proof}[Proof of Proposition~\ref{prop:worstexample}]
Consider any increasing concave function $f$ and a polygon approximation $\tilde{f}_X$ of $f$. Let $X = \{0 = x_0, \cdots, x_k = v_{\max}\}$. Since $f$ is concave, the function $\tilde{f}_X$ lies `below' the function $f$. Consider now the tangents to the function $f$ drawn at points in $X$. Since $f$ is concave, these tangents are always `above' the function $f$. The tangents can be stitched together to define a function $g$ that is always above the function $f$. Formally, define $g(x) = \min_{i \in \{0,\ldots,k\}} \{f(x_i) + \min\{f^-(x_i) \cdot (x-x_i),f^+(x_i)\cdot (x-x_i)\}\}$. Observe that $f(x_i) = g(x_i)$, $f^-(x_i) = g^-(x_i)$, and $f^+(x_i) = g^+(x_i)$ for all $i$, giving us the second bullet. Observe also that $g$ has at most $2k$ different slopes (because $x_0$ and $x_k$ each only contribute one possibility, while the rest contribute two possibilities), so it is piecewise-linear with at most $2k$ segments.

It remains to show that there exists a polygon approximation $\tilde{f}_X$ we can start with so that there is no polygon approximation of $g$ using $k$ segments for additive error $\varepsilon$. Let $X$ be $k+1$ points so as to minimize $\max_{x} \{f(x) - \tilde{f}_X(x)\}$, and call this value $\delta$ (by hypothesis, $\delta > \varepsilon$). This guarantees us the following property on $X$:
$$\max_{x \in [x_i, x_{i+1}]}\{f(x) - \tilde{f}_X(x)\} = \delta \text{ for all }i.$$

If not, then there is some $i$ where $\max_{x \in [x_i, x_{i+1}]}\{f(x) - \tilde{f}_X(x)\} < \delta$. $x_i$ could then be moved a teeny bit to the left (making the approximation within the interval $[x_i,x_{i+1}]$ a teeny bit worse) so that $\max_{x \in [x_i, x_{i+1}]}\{f(x) - \tilde{f}_X(x)\}$ remains $< \delta$, but $\max_{x \in [x_{i-1}, x_{i}]}\{f(x) - \tilde{f}_X(x)\}$ becomes $< \delta$ as well (as it also makes the approximation within the interval $[x_{i-1},x_i]$ a teeny bit better, and it was $\leq \delta$ to begin with). Iterating this procedure would then let us make $\max_{x \in [x_j, x_{j+1}]}\{f(x) - \tilde{f}_X(x)\} < \delta$ for all $j \leq i$. And a similar argument for moving $x_{i+1}$ a teeny bit to the right allows us to claim the same for $j \geq i$. At the end, we have now made $\max_{x} \{f(x) - \tilde{f}_X(x)\}$ strictly less than $\delta$, a contradiction.

Now that we have $\max_{x \in [x_i, x_{i+1}]} \{f(x) - \tilde{f}_X(x)\} = \delta > \varepsilon$ for all $i$, we can immediately see that for any function $g(\cdot)$ with $g(x_i) = f(x_i)$ for all $i$, and $g(x) \geq f(x)$ for all $x$, any $Y$ such that $\tilde{g}_Y(\cdot)$ is an $\varepsilon$-approximation polygon approximation requires at least one point $y \in Y \cap [x_i, x_{i+1}]$. So we can immediately see that $g$ also has no polygon approximation using $k$ segments for additive error $\varepsilon$. 
\end{proof} 

With this in mind, we'll now restrict attention to piecewise-linear functions with not too many segments. To this end, first consider a piecewise-linear function with just two segments. The first segment is of length $l_1$ with slope $m_1$. Similarly, the second segment is of length $l_2$ and slope $m_2 < m_1$. If we wish to approximate this function using just one segment, the segment should have length $l_1 + l_2$ and slope $\frac{m_1l_1 + m_2l_2}{l_1+l_2}$. The error in this approximation will be  $\frac{(m_1 -m_2)l_1l_2}{l_1+l_2}$. This calculation says that the error roughly depends on the product of the slope difference and the lengths. For our lower bound, we would wish to have many such segments combined together so that the error in all of the segments is the same. For a concave function, the slope keeps decreasing as $x$ grows. Thus, if the error has to be the same in every segment, the lengths of each segment have to keep increasing so that the product is the same. Observe that the function $\log (x)$ satisfies this property. 

With this direction in mind, we now demonstrate a function that we call $LPL_k$ (for logarithmically-piecewise-linear) that can't be approximated using less than $k$ segments. We chose this name because the function has derivative $1/2^i$ in the range $[3(2^{i+1} - 2), 3(2^{i+2} - 2)]$ which is (roughly) a constant factor away from the derivative $1/(x + 6)$ of the logarithmic function $\ln(x+6)$. In fact, what we do can be seen as stitching together appropriate tangents to the logarithmic curve. For a value of $\varepsilon < 1/100$, let $k = \frac{1}{100\varepsilon}$. Consider the following function defined on $[0,3(2^{k+1}-2) = v_{\max}]$:

\[LPL_k (x) = \begin{cases} x, & 0\leq x\leq 3(2^2-2)\\
 \frac{1}{2}(x -3(2^2-2))  + 6, & 3(2^2-2)<x\leq 3(2^3-2)\\
  \frac{1}{4}(x -3(2^3-2))  + 12, & 3(2^3-2)< x\leq 3(2^4-2)\\
  \ \ \ \ \vdots&\\
   \frac{1}{2^{k-1}}(x -3(2^{k}-2))  + 6(k-1), & 3(2^{k}-2)< x\leq 3(2^{k+1}-2).\\
\end{cases}\]

The function $LPL_k$ satisfies $\varepsilon (1 + LPL_k(v_{\max})) = \varepsilon + 6k\varepsilon \leq  \frac{1}{2}$, and that $f'(0) \leq 1$, $f'(v_{\max}) \geq 0$. We prove that any good polygon approximation for $LPL_k(x)$ must have at least $\Omega(k)$ segments. Since $k = \frac{1}{100\varepsilon}$, this proves that a dependence of $1/\varepsilon$ is unavoidable if we want to be independent of $v_{\max}$. Since in this example, $\sqrt{\frac{\log v_{\max}}{\varepsilon}} = \Theta(k)$, this also shows that the bound in \autoref{thm:polygonapproximation} is tight.  The idea of our proof is to identify disjoint intervals such that any sequence $X$ that defines a good polygon approximation should have at least one point in every such interval. We will identify $\Theta(k)$ such intervals for $LPL_k$. 

\begin{lemma} \label{lemma:LPL} Suppose there exists a sequence $X = \{x_i\}_{i\geq 1}$ of points such that $LPL_k(x) - 1/2 \leq \widetilde{LPL_k}_X(x) \leq LPL_k(x)$ for all $x\in [0,v_{\max}]$. The sequence $X$ has at least one point in the range $I_i = [3(2^i-2)-2^i, 3(2^i-2)+2^i]$ for all $2 \leq i \leq k$.
\end{lemma}

\begin{proof}[Proof of \autoref{lemma:LPL}] 
Suppose for the sake of contradiction there is an $i$ for which $X\cup I_i = \phi$. We prove that the approximation doesn't hold at the midpoint of $I_i$ (call this $m$). Since no point of $X$ is in $I_i$ and $LPL_k$ is concave, $\tilde{LPL_k}_X(m)$ is at most $\frac{1}{2}(LPL_k(3(2^i-2)-2^i) + LPL_k(3(2^i-2)+2^i))$. Thus,
$LPL_k(m) - \tilde{LPL_k}_X(m) \geq 6(i-1) - \frac{1}{2}\left(6(i-2) + 2 +  6(i-1) + 2\right) = 1 .$
\end{proof}

The following corollary follows immediately from Lemma~\ref{lemma:LPL}.

\begin{corollary}\label{cor:main} An $\varepsilon (1+LPL_k(v_{\max}))$ additive polygon approximation of $LPL_k$ above requires at least $k-1$ segments.
\end{corollary}

\autoref{cor:main} shows that the bounds of \autoref{thm:polygonapproximation} are tight, both in the case that there is no dependence on $v_{\max}$, and in the case that we allow dependence $v_{\max}$. Moreover, \autoref{prop:worstexample} shows that ``LPL-like'' functions exhibit the worst-case gaps for whatever kinds of guarantees are desired. 

\subsection{From Polygon Approximation to Approximate Auctions} \label{sec:approxauction}

In this subsection we prove \autoref{prop:curves} and \autoref{cor:curves} establishing the connection between "small" polygon approximations and approximate auctions with low menu complexity. The high level idea of the proof of \autoref{prop:curves} is that the revenue generated by a mechanism on days $i$ through $n$ is equal to a weighted sum of the function $R_{\geq i}(\cdot)$ evaluated at points where the allocation curve changes (see \autoref{lem:geqi}). If  a good polygon approximation of $R_{\geq i}$ exists, then every point where the allocation curve changes can be substituted by the endpoints of the segment of polygon approximation that contains it. The error in the polygon approximation would become loss in revenue after this transformation. The number of segments (or endpoints) in the polygon approximation would be the menu complexity of the transformed mechanism. Thus, a good and small polygon approximation would generate an almost optimal and low menu complexity mechanism.
%The first proof relies on the existence of good, small polygon approximations to create menus that can be turned into incentive compatible ones by at most doubling the number of options offered on a given deadline. The guarantees on the quality of the polygon approximation ensure that the resulting mechanism will recover a large fraction of the optimal revenue.
The proof of \autoref{cor:curves} relies on iteratively calling on \autoref{prop:curves} from the first day to the last while keeping a check on the total revenue lost compared to the optimal over all days.  

\begin{proposition}\label{prop:curves}
Consider a FedEx instance with $n$ deadlines. For $i \in \{1,2,\cdots,n\}$, let $g$ be the function $\tilde{R}_{\geq i}$ defined in \autoref{def:revcurves}, and let $X$ be a sequence of $k$ points in $[0,r_{\geq i}]$ such that for all $x \leq r_{\geq i}$, we have $g(x) - \varepsilon \leq  \tilde{g}_X(x) \leq  g(x) $. Then for any mechanism $M$, there exists a mechanism with the following properties: 

\begin{itemize}
\item The menu offered on days $1,2,\cdots, i-1$ is the same as $M$.
\item The $i$-deadline menu complexity is at most $2k$.
\item The revenue is at least $\mathsf{OPT}_{M,i}-\varepsilon$.
\end{itemize}

Here, $\mathsf{OPT}_{M,i}$ denotes the optimal revenue for any mechanism that offers the same menu as $M$ on deadlines $1,2,\cdots, i-1$.
\end{proposition}

Intuitively, every segment of $\tilde{g}_X$ corresponds to a different menu option in the approximate mechanism, and the fact that $\tilde{g}_X(x)$ and $g(x) = \tilde{R}_{\geq i}(x)$ are close for all $x$ implies that the revenue of the mechanism based on $\tilde{g}_X$ isn't far from optimal (by \autoref{lem:geqi}). The additional factor of two comes from the fact that $\tilde{R}_{\geq i}$ is ironed, so to ``set price $p$'' in $\tilde{R}_{\geq i}$, we might need to randomize over two prices in $R_{\geq i}$.

%\begin{proof}[Proof of \autoref{cor:curves}] The proof of \autoref{prop:curves} defines an approximate auction where the approximation is performed on day $i$. The total revenue lost on subsequent days is at most $\varepsilon$. Even though the statement there is expressed in term of the optimal auction, we only use the fact that mechanism behaves optimally on day $i+1$ onwards. In this light, we can invoke \autoref{prop:curves} $n$ times, starting from $i=1$ and $G_i$, each time losing $\frac{\epsilon}{n}R_{\geq i}(r_{\geq i}) \leq \frac{\epsilon}{n}\mathsf{OPT}$. The total loss in revenue is at most $\epsilon \mathsf{OPT}$, and the $i$ deadline menu complexity is $k_i$ for all days $i$. Thus the menu complexity is at most $2 \sum_i k_i$.
%\end{proof}

\begin{proof}[Proof of \autoref{prop:curves}] We assume without loss of generality that the mechanism $M$ does the optimal pricing on days $i$ through $n$ (given the prices it sets on days $1$ through $i-1$). Thus, the revenue generated by $M$ is $\mathsf{OPT}_{M,i}$. The menu offered by any such mechanism on day $i$ can be seen as a distribution $f$ over prices $\{p_j\}$. The prices $p_j$ are all at most $r_{\geq i}$ as the $M$ behaves optimally on day $i$ (see \autoref{def:optcurves}). Since the mechanism behaves optimally day $i+1$ onwards, the revenue generated by $M$ on days $i$ through $n$ is $\sum_j f(p_j) R_{\geq i} (p_j)$ by \autoref{lem:geqi}.

Let $X$ be the sequence of $k$ points such that $g(x) - \varepsilon \leq  \tilde{g}_X(x) \leq  g(x) $ for all $x \leq r_{\geq i}$. Since $g  = \tilde{R}_{\geq i}$ is the upper concave envelope of $R_{\geq i}$, for every point $x \leq r_{\geq i}$, there exists points $x_l \leq x$ and $x_r \geq x$ in $[0,r_{\geq i}]$ such that $g(x_l) = R_{\geq i}(x_l)$,  $g(x_r) = R_{\geq i}(x_r)$, and for any $\lambda \in [0,1]$, we have $g\left(\lambda x_l + (1-\lambda)x_r\right) = \lambda g(x_l) + (1-\lambda)g(x_r)$. For all points $x \in X$, we add the points $x_l$ and $x_r$ to $X$ and remove the point $x$.  Adding the points $x_l$ and $x_r$ will only improve the quality of the polygon approximation. Further, since $g$ is linear on the interval $[x_l, x_r]$ that contains $x$, removing $x$ doesn't affect the quality of the polygon approximation. Thus, the new set $X$ satisfies

\begin{equation} \label{eq:newsetx1}
g(x) - \varepsilon \leq  \tilde{g}_X(x) \leq  g(x).
\end{equation}
Further, by construction all elements $v \in X$ satisfy

\begin{equation} \label{eq:newsetx2}
g(v) = R_{\geq i}(v). 
\end{equation}

The set $X$ now has at most $2k$ points. For any $x \leq r_{\geq i}$, we let $\underline{x}$ (resp. $\overline{x}$) denote the largest (resp. smallest) value in $X$ that is at most (at least) $x$. With this notation, if $x = \lambda_x \underline{x} + (1-\lambda_x)\overline{x}$, then we have that

\begin{equation} \label{eq:interpolate}
\tilde{g}_X(x) = \lambda_x \tilde{g}_X(\underline{x}) + (1-\lambda_x)\tilde{g}_X(\overline{x}) = \lambda_x g(\underline{x}) + (1-\lambda_x)g(\overline{x}).
\end{equation}

We are now ready to define a mechanism $M'$ with high revenue and low $i$-deadline menu complexity. The mechanism $M'$ is defined as follows:

\begin{itemize}
\item Mimic $M$ on days $1,2,\cdots, i-1$.
\item On day $i$, if $p_j$ has mass $f(p_j)$, add mass $f(p_j) \lambda_{p_j}$ to $ \underline{p_j}$ and mass $f(p_j)(1- \lambda_{p_j})$ to $ \overline{p_j}$. This generates a new distribution $f'$ on $X$ which defines our menu for day $i$.
\item Do the optimal pricing days $i+1$ onwards given the prices on day $1$ through $i$. 
\end{itemize}

Since the menu offered on day $i$ is defined by a distribution over $X$, it's size is at most $2k$, the size of $X$. We finish the proof by showing that $M'$ is feasible and has high revenue. 

\begin{claim} The mechanism $M'$ defined above is feasible.
\end{claim}
\begin{proof} We only need to check the intra-day incentive compatibility constraints. For days $1$ through $i-1$, $M'$ mimics $M$ and thus, these constraints are satisfied. Days $i+1$ onwards, $M'$ prices optimally and thus, satisfies the constraints. We only argue about the constraints for day $i$. We wish to prove that the utility of any bidder with type $(x,i)$ in mechanism $M'$ is at least that of the bidder with type $(x, i-1)$. We know that  the utility of any bidder with type $(x,i)$ in mechanism $M$ is at least that of the bidder with type $(x, i-1)$. Since the mechanisms $M$ and $M'$ are the same on days $1$ through $i-1$, it is sufficient to prove that the utility of any bidder with type $(x,i)$ in mechanism $M'$ is at least that of the bidder with type $(x, i)$ in mechanism $M$. This follows directly from the expression of utilitiy $\sum_{p_j \leq x} f(p_j)(x - p_j)$. We have 

\begin{align*}
&\sum_{X \ni v\leq x} f'(v)(x-v) \\
&=\sum_{p_j \leq \underline{x}} f(p_j) \lambda_{p_j} (x - \underline{p_j}) +  f(p_j)(1 - \lambda_{p_j}) (x - \overline{p_j})  + \sum_{ p_j \in (\underline{x},x]} f(p_j)\lambda_{p_j} (x - \underline{p_j})&\text{(Definition of $f'$)} \\
&= \sum_{p_j \leq \underline{x}} f(p_j) (x -p_j) +  \sum_{ p_j \in (\underline{x},x]} f(p_j)\bigg(1 - \frac{p_j-\underline{p_j}}{\overline{p_j}-\underline{p_j}}\bigg) (x - \underline{p_j})&\text{(Definition of $\lambda_{p_j}$)} \\
&\geq \sum_{p_j \leq \underline{x}} f(p_j) (x -p_j) +  \sum_{ p_j \in (\underline{x},x]} f(p_j)\bigg(1 - \frac{p_j-\underline{p_j}}{x-\underline{p_j}}\bigg) (x - \underline{p_j}) &\text{(Since  $\overline{p_j} > x$)} \\
&=  \sum_{p_j \leq \underline{x}} f(p_j) (x -p_j) +  \sum_{ p_j \in (\underline{x},x]} f(p_j) (x - p_j)  =  \sum_{p_j \leq x} f(p_j)(x-p_j).\\
\end{align*}
\end{proof}

\begin{claim} The revenue generated by $M'$ is at least $\mathsf{OPT}_{M,i}-\varepsilon$.
\end{claim}
\begin{proof} First, we note $M$ and $M'$ are identical for the first $i-1$ days. Thus, the revenue generated on the first $i-1$ days is the same for both $M$ and $M'$. Since $M'$ prices optimally day $i+1$ onwards, the revenue generated on day $i$ through $n$ is $\sum_{v \in X} f'(v) R_{\geq i}(v)$. We have 
\begin{align*}
\sum_{v \in X} f'(v) R_{\geq i}(v)& = \sum_{v \in X} f'(v) g(v) &\text{(Equation~\ref{eq:newsetx2})}  \\
&=\sum_{v_j} f(v_j) \lambda_{v_j} g(\underline{v_j}) + f(v_j) (1- \lambda_{v_j}) g(\overline{v_j}) &\text{(Definition of $f'$)} \\
&=\sum_{v_j} f(v_j) \tilde{g}_X(v_j) &\text{(Equation~\ref{eq:interpolate})} \\
&\geq \sum_{v_j} f(v_j) \left(g(v_j) - \varepsilon\right) &\text{(Equation~\ref{eq:newsetx1})} \\
&\geq \sum_{v_j} f(v_j) R_{\geq i}(v_j) -\varepsilon.  \\
\end{align*}

This analysis proves that the total revenue generated by  $M'$ is at most $\varepsilon$ less than that generated by $M$. In other words, it is at least $\mathsf{OPT}_{M,i}-\varepsilon$.

\end{proof}
These two claims combined suffice to prove the proposition.
\end{proof}

Having described the construction of low $i$-deadline menu complexity mechanisms, we repeat this construction $n$ times (once for each deadline) to get our approximately optimal mechanism. We also show how to use \autoref{prop:curves} to go from the optimal mechanism to an entire menu of low menu complexity. Thus, the total revenue loss on days $i$ through $n$ is at most $\varepsilon$. Using the above result, we prove Corollary~\ref{cor:curves}.

\begin{proof}[Proof of Corollary~\ref{cor:curves}] We prove that for all $i \in \{0,1,\cdots, n\}$, there exists a mechanism with $j$-deadline menu complexity of $k_j$ for all $1\leq j \leq i$ whose revenue is at least $\mathsf{OPT} - \sum_{j=1}^i \varepsilon_j$. For $i=n$, this is the same as the statement of the corollary. This proof will proceed via induction on $i$. For $i=0$, the statement is trivial. We assume the statement for $i-1$ and prove it for $i$. 

Let $M_{i-1}$ be the mechanism that is promised by the induction hypothesis and let $\mathsf{Rev}_{i-1} \geq \mathsf{OPT} - \sum_{j=1}^{i-1} \varepsilon_i$ be its revenue. 
%Define $\mathsf{OPT}_{i-1}$ to be the maximum revenue of mechanism that matches $M_{i-1}$ on days $1,2,\cdots, i-1$. We get 

%\[\mathsf{OPT}_{i-1} \geq \mathsf{Rev}_{i-1}  \geq \mathsf{OPT} - \sum_{j=1}^{i-1} \varepsilon_i.\]

We invoke Proposition~\ref{prop:curves} on $X_i$ and $M_{i-1}$ to get a mechanism $M_i$ with the following properties:

\begin{itemize}
\item The menu offered by $M_i$ on days $1,2,\cdots, i-1$ is the same as $M_{i-1}$.
\item The $i$-deadline menu complexity is at most $2k_i$.
\item The revenue is at least $\mathsf{OPT}_{M_{i-1},i}-\varepsilon_i$.
\end{itemize}

Property $1$ and $2$ imply that the $j$-deadline menu complexity of $M_i$ is at most $k_j$ for all $1\leq j \leq i$. The revenue of $M_i$ is at least $\mathsf{OPT}_{M_{i-1},i}-\varepsilon_i \geq \mathsf{Rev}_{i-1}-  \varepsilon_i \geq \mathsf{OPT} - \sum_{j=1}^{i-1} \varepsilon_j- \varepsilon_i =  \mathsf{OPT} - \sum_{j=1}^{i} \varepsilon_j$. This completes the proof.

\end{proof}

Finally, we may complete the proof of~\autoref{thm:ubmain}.

\begin{proof}[Proof of \autoref{thm:ubmain}]
For any FedEx instance where the bidder's types have integral support $1,2,\cdots ,v_{\max}$, observe that the optimal revenue is at least $1$. This is because $1$ is the revenue generated by the auction that offers the price $1$ on all deadlines. Further, for all deadlines $i$, $1+ \tilde{R}_{\geq i}(v_{\max}) \leq \mathsf{OPT} + \mathsf{OPT} = 2\mathsf{OPT}$.

Note that all revenue curves $\tilde{R}_{\geq i} = g_i$ when restricted to $[0, r_{\geq i}]$ satisfy the requirements of \autoref{thm:polygonapproximation}. This implies that for all $i$, there exists sequences $X_i$ of at most $O\left(\min\left\{n/\varepsilon,\sqrt{\frac{n}{\varepsilon}\log v_{\max}}\right\}\right)$ points such that for all $x \in [0,r_{\geq i}],$
\[g_i(x)-\frac{\varepsilon}{2n} \left(1+g_i(v_{\max})\right) \leq \tilde{g_i}_{X_i}(x) \leq g_i(x).\]

We set $\varepsilon_i = \frac{\varepsilon}{2n} \left(1+g_i(v_{\max})\right)$ and use \autoref{cor:curves} to get that there exists a mechanism with menu complexity at most $O\left(n\min\left\{n/\varepsilon,\sqrt{\frac{n}{\varepsilon}\log v_{\max}}\right\}\right)$ whose revenue is at least $ \mathsf{OPT} - \sum_{i=1}^n\varepsilon_i  = \mathsf{OPT} - \sum_{i=1}^n\frac{\varepsilon}{2n} \left(1+g_i(v_{\max})\right) \geq \mathsf{OPT} - \sum_{i=1}^n\varepsilon\frac{\mathsf{OPT}}{n} = (1-\varepsilon)\mathsf{OPT}$.

%Define the functions $g_i$ to be the $\frac{\varepsilon}{2n}(1+ \tilde{R}_{\geq i}(v_{\max}) \leq \frac{\varepsilon \mathsf{OPT}}{n}$ additive polygon approximation to the functions $\tilde{R}_{\geq i}(\cdot)$. Since all the revenue curves $\tilde{R}_{\geq i}$ satisfy the requirements of \autoref{thm:polygonapproximation}, we get that $g_i$ can be constructed using at most $k_i = O\left(\min\left\{n/\varepsilon,\sqrt{\frac{n}{\varepsilon}\log v_{\max}}\right\}\right)$ segments. We then apply \autoref{cor:curves} on $g_i$ and $k_i$ to get that there exists a mechanism that generates at least $\mathsf{OPT} - \varepsilon \mathsf{OPT}$ whose menu complexity is at most $\sum_{i=1}^n k_i = O\left(n\min\left\{n/\varepsilon,\sqrt{\frac{n}{\varepsilon}\log v_{\max}}\right\}\right)$.
%Observe first that $\mathsf{OPT} \geq R_{\geq i}(r_{\geq i})$ for all $i$ (and in fact $\mathsf{OPT} = R_{\geq 1}(r_{\geq 1})$. Moreover, observe that as all distributions have integral support in $\{1,v_{\max}\}$, we also have $\mathsf{OPT} \geq 1$. Therefore, we get that $R_{\geq i}(r_{\geq i}) + 1 \leq 2 \mathsf{OPT}$, and the guarantees provided by Theorem~\ref{thm:polygonapproximation} imply the bounds required by Corollary~\ref{cor:curves}.
\end{proof}

\section{Analysis omitted in Section~\ref{sec:LBA}}
\label{app:LBA}
\subsection{The instance}  \label{sec:approxinstance}

In this section, we describe our instance of the FedEx problem that is hard to approximate using small menus. Before delving into the details of our parameters, we give a brief high level intuition.

In Section~\ref{sec:tightpolygon}, we described a function $LPL_k$ that is hard to polygon approximate. The function $LPL_k$ provided a lower bound for Theorem~\ref{thm:polygonapproximation} that was tight (modulo constant factors).  The idea behind this example was to have a piecewise linear function such that the slope of the constituent segments decreases gradually in a controlled way.  Because of the intimate connection between polygon approximation and menu complexity (Proposition~\ref{prop:curves}), similar ideas should also help in constructing a FedEx instance that is hard to approximate using small menus.

The construction of the hard FedEx instance can be expected to have many additional complications. The complications arise from the fact that each revenue curve $R_{\geq i}$ in a FedEx instance is formed by appropriately adding revenue curves of multiple distributions (see Definition~\ref{def:revcurves}) and $\tilde{R}_{\geq i}$ is obtained by `ironing' $R_{\geq i}$. Since the size of the optimal menu only increases when prices `split' across ironed intervals, it is necessary to have multiple (polynomial, in our case) ironed intervals in the function $\tilde{R}_{\geq i}$. 

The presence of multiple ironed intervals in $\tilde{R}_{\geq i}$ is, however, not sufficient. It is easy to see why. Imagine an example where all the revenue curves $R_{i}$ are generated from the same underlying distribution. If this is the case, all the curves $R_{\geq i}$ might have the same ironed intervals.\footnote{This is not always true as whether a particular interval is ironed or not for a given deadline depends on many other factors, {\em e.g.} the distribution $q_i$ across the various days.} If the ironed intervals in the curves $\tilde{R}_{\geq i}$ for all $i$ correspond, then even the optimal revenue auction would require only $1$ price on each day. This is because only those prices `split' that are inside an ironed interval. If all the ironed intervals correspond, then no price can ever lie inside a interval. The solution is to have ironed intervals `nested' so that the ends of the interval on day $i$ lie inside the intervals on day $i+1$. 

The two points raised above concern splitting in general and were also true of our worst case example described in Section~\ref{sec:LB}. If we want a lower bound for all approximate auctions, we also want the slope between two consecutive segments to be very different (as in $LPL_k$). This ensures that we need at least one menu option for each segment for a good approximation.

With these ideas in mind, we describe our hard instance. We assume that the bidder's type is drawn from a distribution supported on $\{1,2,\cdots, v_{\max}\}\times\{1,2,\cdots, n\}$. The value of $v_{\max}$ in our example is $5n$ and the distribution $q$ over days is uniform. We make explicit that we don't include the $q$ in our calculation. Since they only scale all revenue curves by $1/n$, they don't affect the quality of a $1-\epsilon$ approximation.  The marginal distribution $F_i$ on day $i$ is designed so that the revenue curve $\tilde{R}_{\geq i}$ has $i+1$ segments. The slope of the first $i$ segments is geometrically decreasing with common ratio $\lambda = 1+ \frac{1}{4n}$. This value is large enough so that a $1-O(1/n^2)$ approximation will require one menu option for each segment. It also satisfies, 

\begin{claim} \label{claim:lambdapowers} For $n \geq i \geq 0$, $\frac{1}{\lambda^{i}} > \frac{3}{4}$.
\end{claim}
\begin{proof}
\[\frac{3}{4}\lambda^i < \frac{3}{4}\Big(1+\frac{1}{4n}\Big)^n \leq \frac{3}{4}\mathrm{e}^{\frac{1}{4}} < 1.\]
\end{proof}

As we define the slopes to be in geometric progression, the geometric sum $1 + \frac{1}{\lambda} \cdots$ will appear multiple times in our description and its analysis. To avoid writing the all the terms everywhere, we define $S_0 = n $ and $S_i = n + \sum_{j=0}^{i-1} \lambda^{-j}$ for all $i \geq 1$. 

The $(i+1)^{\text{th}}$ segment of $\tilde{R}_{\geq i}$ will have slope $(n+1-i)C$ where $C = 1/2$. This last segment will split into two segments on day $i+1$. The first segment will have slope $1/\lambda^i$, continuing the geometric progression. The last segment will have slope $(n-i)C$, allowing it to be split similarly in the future. In order to ensure this behavior of $\tilde{R}_{\geq i}$, we need the slope of $R_i$ to be $\beta_i$ for this segment, where  
\begin{equation} \label{eq:beta}
\beta_{i} = \frac{1}{3n+i} \left(\frac{3n-i}{2} - \frac{n-i}{\lambda^{i}} + S_i\right).
\end{equation}

This value of $\beta_i$ satisfies the following two properties:

\begin{claim} \label{claim:betamonotone} $\beta_i$ is an increasing sequence for $n \geq i \geq 0$.
\end{claim}
\begin{proof}
Fix $n > i \geq 0$.
\begin{equation*}
\begin{aligned}
&\beta_{i+1} - \beta_i \\
&= \frac{1}{3n+i+1}\left(\frac{3n-i-1}{2} - \frac{n-i-1}{\lambda^{i+1}} + S_{i+1}\right) - \frac{1}{3n+i}\left(\frac{3n-i}{2} - \frac{n-i}{\lambda^{i}} + S_{i}\right) & \text{(\autoref{eq:beta})} \\
&= \frac{1}{(3n+i)(3n+i+1)}\left(-3n + \frac{(\lambda-1)(3n+i+1)(n-i)}{\lambda^{i+1}} + \frac{4n}{\lambda^{i+1}} + \frac{3n+i}{\lambda^{i}} -S_i\right) \\
&\geq \frac{1}{(3n+i)(3n+i+1)}\left(\frac{3(3n+i+1)(n-i)}{16n}+ \frac{3(3n+i)}{4} -S_i\right) & \text{(Claim~\ref{claim:lambdapowers})}\\
&\geq \frac{1}{(3n+i)(3n+i+1)}\left(\frac{9n}{4} -2n\right) > 0. \\
\end{aligned}
\end{equation*}

\end{proof}

\begin{claim} \label{claim:betarange} For $n \geq i \geq 1$, $\frac{1}{2} < \beta_1 = \frac{1}{2} + \frac{5}{4\lambda(3n+1)} \leq \beta_{i} \leq \frac{3}{4}$.
\end{claim}
\begin{proof} Verify that $\beta_{1}$ is indeed correct. After proving Claim~\ref{claim:betamonotone}, it is sufficient to establish $\beta_n \leq \frac{3}{4}$. 

$$\beta_n = \frac{1}{4n}(n+ S_n) \leq  \frac{3n}{4n}.$$
\end{proof} 

We now define the marginal distribution $F_i$ on the  $i^{\text{th}}$ deadline. We write a expression for the distribution and then, calculate the revenue curve $R_i$ of the distribution $F_i$. The part of the analysis where we prove that $F_i$ is a valid distribution is omitted. This would use $\frac{S_i}{n+i} \geq \frac{S_{i+1}}{n+i+1}$ and $\beta_i \leq \frac{S_i}{n+i}$ (where  Claim~\ref{claim:lambdapowers} and Claim~\ref{claim:betarange} help prove the latter) and can easily be done by the interested reader. The distribution over types on the $i^{\text{th}}$ day is given by :
%The revenue curve for this distribution is given by :

%In our hard instance, the bidder's values are positive integers at most $v_{\max} = O(n)$. We make sure that any mechanism that generates at least $1-O(1/n^2)$-fraction of the optimal revenue requires a menu of size at least $\Omega(n)$ on $\Omega(n)$ days. This means that the menu complexity required for a $1-O(1/n^2)$-approximation is $\Omega(n^2)$.  For the purposes of this example, set $\lambda = 1 + 1/4n$, and $S_i = n + \sum_{j=0}^{i-1} \lambda^{-j}$. Assume that $S_0 = n$. Also, define $C=1/2$ and \[\beta_{i} = \frac{1}{3n+i}\Bigg(\frac{3n-i}{2} - \frac{n-i}{\lambda^{i}} + S_i\Bigg).\]

% Let the distribution $q$ over days be uniform. We want to make explicit that we don't include the $q_i$s in our calculation. Since they only scale all revenue curves by $1/n$, they don't affect the quality of a $1-\epsilon$ approximation.
%The distribution over types on the $i^{\text{th}}$ day is given by 

\begin{equation} \label{eq:typedis}
F_{i}(x) = \begin{cases} 
      0, &0 \leq x \leq n \\
      \Big(1-\frac{S_1}{n+1}\Big) ,&n \leq x \leq n+1 \\
      \Big(1-\frac{S_2}{n+2}\Big) ,&n+1\leq x\leq n+2 \\
      \Big(1-\frac{S_3}{n+3}\Big) ,&n+2\leq x\leq n+3 \\
      \ \ \ \ \ \ \ \vdots &\\
      \Big(1-\frac{S_{i}}{n+i}\Big),&n+i-1\leq x\leq n+i \\
      (1-\beta_{i}) ,&n+i\leq x\leq 3n+i \\
      1,&3n + i\leq x \leq 5n. \\
   \end{cases}
   \end{equation}

%To verify that this is a valid distribution, see that $\frac{S_i}{n+i} \geq \frac{S_{i+1}}{n+i+1}$ and $\beta_i \leq \frac{S_i}{n+i}$ (using Claim~\ref{claim:lambdapowers} and Claim~\ref{claim:betarange}  for the latter). 
The revenue curve for this distribution is given by $R_i(v) = v\left(1-F_i(v)\right) $:

\begin{equation} \label{eq:ri}
R_{i}(x) = \begin{cases} 
      x, &0 \leq x \leq n \\
      \left(\frac{S_1}{n+1}\right)x ,&n < x \leq n+1 \\
      \left(\frac{S_2}{n+2}\right)x ,&n+1< x\leq n+2 \\
      \left(\frac{S_3}{n+3}\right)x ,&n+2< x\leq n+3 \\
      \ \ \ \ \ \ \ \vdots &\\
      \left(\frac{S_{i}}{n+i}\right)x,&n+i-1< x\leq n+i \\
      \beta_{i}x ,&n+i< x\leq 3n+i \\
      0,&3n+i< x \leq 5n. \\
   \end{cases}
\end{equation}

Using Definition~\ref{def:revcurves}, we now calculate the combined revenue curve $R_{\geq i}$ for days $i$ through $n$.  We prove that $R_{\geq i}$ has the form we desire using induction. 
%Let $\tilde{R}_{\geq i}$ denote the combined ironed revenue curve for days $i$ through $n$ (both inclusive). Let $R_{\geq i}$ denote the version before ironing. We prove by induction that 
\begin{theorem} \label{thm:revcurvesapprox}
\begin{equation} \label{eq:rgeqi}
R_{\geq i}(x)= \begin{cases} 
      x + (n-i)x, &0 \leq x \leq n \\
      \left(\frac{S_1}{n+1}\right)x + (n-i)\left(x-n + S_0\right), &n < x \leq n+1 \\
      \left(\frac{S_2}{n+2}\right)x + (n-i)\left(\frac{x-n-1}{\lambda}+S_1\right), &n+1 < x\leq n+2 \\
      \left(\frac{S_3}{n+3}\right)x + (n-i)\left(\frac{x-n-2}{\lambda^{2}}+S_2\right), &n+2 < x\leq n+3 \\
      \ \ \ \ \ \ \ \vdots &\\
      \left(\frac{S_{i}}{n+i}\right)x + (n-i)\left(\frac{x+1-n-i}{\lambda^{i-1}} + S_{i-1} \right), &n+i-1 < x\leq n+i \\
      \beta_{i}x + (n-i)\left(\frac{x-n-i}{\lambda^{i}}+S_{i} \right),&n+i < x\leq n+i+1 \\
      \beta_{i}x + (n-i)\left(C(x-n-i-1)+S_{i+1} \right),&n+i+1 < x\leq 3n+i \\
      (n-i)\left(C(x-n-i-1)+S_{i+1} \right), &3n+i\ < x\leq 3n+i+1 \\
      (n-i)\left(2nC+S_{i+1}\right),&3n+i+1 <  x \leq 5n. \\
   \end{cases}
\end{equation}
\end{theorem}

\begin{proof} Proof by backwards induction. The base case $i = n$ is easily verified. For the inductive step, we first calculate $\tilde{R}_{\geq i}$. This is done by ironing the curve $R_{\geq i}$ defined in \autoref{eq:rgeqi}. We provide an expression for $\tilde{R}_{\geq i}$ and prove that it is correct in Lemma~\ref{lemma:ironingLBAexample} in Subsection~\ref{app:ironingLBAexample} of this Appendix. The expression is:

\begin{equation} \label{eq:tildergeqi}
\tilde{R}_{\geq i}(x) = \begin{cases} 
      (n+1-i)x, &0 \leq x \leq n \\
      (n+1-i)(x-n + S_0)  ,&n < x \leq n+1 \\
      (n+1-i)\left(\frac{x-n-1}{\lambda} +S_1\right) ,&n+1 < x\leq n+2 \\
      (n+1-i)\left(\frac{x-n-2}{\lambda^2}+S_2\right),&n+2 < x\leq n+3 \\
      \ \ \ \ \ \ \ \vdots &\\
      (n+1-i)\left(\frac{x+1-n-i}{\lambda^{i-1}}+S_{i-1} \right),&n+i-1 < x\leq n+i \\
      (n+1-i)\left(C(x-n-i)+S_i \right),&n+i < x\leq 3n+i \\
      (n-i)\left(2nC + S_{i+1}\right) + \frac{(n-i)C - \beta_{i}(3n+i)}{2n -i} (x-5n),&3n + i < x \leq 5n. \\
   \end{cases}
\end{equation}

Note that this function is maximized at $r_{\geq i} = 3n+i$. We now use Definition~\ref{def:revcurves} which says:

\[ R_{\geq i-1}(v) = \begin{cases}
      R_{i-1}(v) + \tilde{R}_{\geq i} (v), & v < r_{\geq i} \\
      R_{i-1}(v) + \tilde{R}_{\geq i} (r_{\geq i}), & v \geq r_{\geq i}. \\
\end{cases} \]

Using  \autoref{eq:ri} and \autoref{eq:tildergeqi} in this gives:
\[R_{\geq i-1}(x) = \begin{cases} 
      x + (n+1-i)x, &0 \leq x \leq n \\
      \left(\frac{S_1}{n+1}\right)x + (n+1-i)(x-n + S_0)  ,&n < x \leq n+1 \\
      \left(\frac{S_2}{n+2}\right)x + (n+1-i)\left(\frac{x-n-1}{\lambda} +S_1\right) ,&n+1 < x\leq n+2 \\
      \left(\frac{S_3}{n+3}\right)x + (n+1-i)\left(\frac{x-n-2}{\lambda^2}+S_2\right),&n+2 < x\leq n+3 \\
      \ \ \ \ \ \ \ \vdots &\\
      \left(\frac{S_{i-1}}{n+i-1}\right)x + (n+1-i)\left(\frac{x+2-n-i}{\lambda^{i-2}}+S_{i-2} \right),&n+i-2 < x\leq n+i-1 \\
      \beta_{i-1}x + (n+1-i)\left(\frac{x+1-n-i}{\lambda^{i-1}}+S_{i-1} \right),&n+i-1 < x\leq n+i \\
      \beta_{i-1}x + (n+1-i)\left(C(x-n-i)+S_i \right),&n+i < x\leq 3n+i-1 \\
      (n+1-i)\left(C(x-n-i)+S_i \right),&3n+i-1 < x\leq 3n+i \\
      (n+1-i)\left(2nC + S_i\right),&3n + i < x \leq 5n. \\
   \end{cases}\]
   
which is of the form required by the induction hypothesis.
\end{proof}

It is relatively easy to find the optimal auction in our instance using  Fiat et al.'s algorithm \cite{Fiat:2016}. The optimal auction closely follows the behavior of the curves $\tilde{R}_{\geq i}$. It has one menu option $= 3n+1$ on day $1$. On each subsequent day, the last menu option splits into two options while all the other options are carried as is. For example, the price $3n+1$ splits into two prices of $n+2$ and $3n+2$ on day $2$. The price $n+2$ is pushed forward to all of the days while the price $3n+2$ is split into $n+3$ and $3n+3$. The price $n+3$ is pushed through while the price $3n+3$ is again spit into two on day $4$. This process goes on till the $n^{\text{th}}$ day. The menu offered on day $n$ has options $n+2, n+3, \cdots, n+n$ and $4n$.

\begin{remark} The optimal revenue in the setting described is at most $\max R_{\geq 1} = n(2n+1)$ which is between $2n^2$ and $3n^2$. This is denoted by $\mathsf{OPT}$ throughout.
\end{remark}

\subsection{Omitted details from the proof of \autoref{thm:revcurvesapprox}} \label{app:ironingLBAexample}

In this subsection we show that $\tilde{R}_{\geq i}(x)$ is the upper concave envelope of $R_{\geq i}(x)$, where both functions are as defined in \autoref{app:LBA}. The proof consists of showing that the two functions agree on the points where $\tilde{R}_{\geq i}(x)$ changes, and that $\tilde{R}_{\geq i}(x)$ is greater than $R_{\geq i}(x)$ on the other points. This, combined with the fact that  $\tilde{R}_{\geq i}(x)$ is linear in between the points where it changes, suffices to show the claim. 	 

\begin{lemma} \label{lemma:verticesmatch} For all $x \in \{0, n, n+1, n+2, \cdots, n+i -1, n+i , 3n+i,5n\}$, we have $\tilde{R}_{\geq i}(x)  =  R_{\geq i}(x)$.
\end{lemma}
\begin{proof}

 We prove this lemma using four claims, all of which establish $\tilde{R}_{\geq i}(x) = R_{\geq i}(x)$ for different domains of $x$.

 \begin{claim} For all $x \in \{0, n\}$, we have $\tilde{R}_{\geq i}(x)  =  R_{\geq i}(x)$.
 \end{claim}
 \begin{proof}
 By verification: $x + (n-i) x = (n+1 -i)x$.
 \end{proof}

 \begin{claim} For all $x \in \{ n+1, n+2, \cdots, n+i -1, n+i\}$, we have $\tilde{R}_{\geq i}(x)  =  R_{\geq i}(x)$.
 \end{claim}
 \begin{proof}
 By verification:
\[\tilde{R}_{\geq i}(x) - R_{\geq i}(x) = \left(\frac{1}{\lambda^{x-n-1}}+S_{x-n-1} \right) - S_{x-n} = 0.\]
 \end{proof}

 \begin{claim}  $\tilde{R}_{\geq i}(3n+i)  =  R_{\geq i}(3n+i)$.
 \end{claim}
 \begin{proof}
%There exists a $1 \leq j \leq i$ such that $x$ satisfies $n + j - 1 < x \leq n+j$. We use this $j$ to evaluate the functions $\tilde{R}_{\geq i}$ and $R_{\geq i}$ at $x$. We get

\begin{align*}
R_{\geq i}(3n+i) &= \beta_{i}(3n+i) + (n-i)\left(C(2n-1)+S_{i+1} \right) \\
&= \left(\frac{3n-i}{2} - \frac{n-i}{\lambda^{i}} + S_i\right) + (n-i)\left(C(2n-1)+S_{i+1}\right) & \text{(\autoref{eq:beta})}\\
&= n + \left((n-i)C - \frac{n-i}{\lambda^{i}} + S_i\right) + (n-i)\left(C(2n-1)+S_{i+1}\right) \\
&= n + S_i + (n-i)\left(2nC+S_{i}\right) \\
&= (n+1-i)\left(2nC+S_i \right) = \tilde{R}_{\geq i}(3n+i). \\
\end{align*}
 \end{proof}

 \begin{claim} $\tilde{R}_{\geq i}(5n) = R_{\geq i}(5n)$.
 \end{claim}
 \begin{proof}
This is easily verifiable from the description of the functions. 
\end{proof}
\end{proof}

\begin{lemma} \label{lemma:tildergeqigeqrgeqi} For all $0 \leq x \leq 5n$, we have $\tilde{R}_{\geq i}(x) \geq R_{\geq i}(x)$.
\end{lemma}
\begin{proof}

Note that both functions are piecewise linear. The function $\tilde{R}_{\geq i}$ changes form at points $S_1 =\{0, n, n+1, n+2, \cdots, n+i -1, n+i , 3n+i,5n\}$. The function $R_{\geq i}$ changes form at points in $S_2 = S_1 \cup \{n+i+1, 3n+i+1\}$. To prove this lemma, it is sufficient to show that  $\tilde{R}_{\geq i}(x) \geq R_{\geq i}(x)$ for all $x \in S_2$. If $x \in S_1$, this follows by \autoref{lemma:verticesmatch}. We only prove for $x  \in \{n+i+1, 3n+i+1\}$.

If $x = n+i+1$, we get
\begin{align*}
\tilde{R}_{\geq i}(n+i+1) & - R_{\geq i}(n+i+1)\\
&= (n+1-i)\left(C+S_i \right)  - \beta_{i}(n+i+1) - (n-i)\left(\frac{1}{\lambda^{i}}+S_{i} \right)\\
&= C+S_i  + (n-i)\left(C-\frac{1}{\lambda^i} \right)  - \beta_{i}(n+i+1)\\
&= C + \beta_i(3n+i) - 2nC - \beta_{i}(n+i+1)&\text{(\autoref{eq:beta})}\\
&= (\beta_i - C)(2n-1) \\
&> 0. &\text{(\autoref{claim:betarange})}\\
\end{align*}

If $x = 3n+i+1$, we get
\[\tilde{R}_{\geq i}(3n+i+1) \geq (n-i)\left(2nC + S_{i+1}\right)  = R_{\geq i}(3n+i+1),\]
as the other term in the definition of $\tilde{R}_{\geq i}$ is positive.

\end{proof} 
 
\begin{lemma} \label{lemma:ironingLBAexample} The upper concave envelope of the function $R_{\geq i}$ defined in \autoref{eq:rgeqi} is the function $\tilde{R}_{\geq i}$ defined in \autoref{eq:tildergeqi}.
\end{lemma}
\begin{proof}
By definition, $\tilde{R}_{\geq i}$ is a continuous piecewise linear function whose successive segments have decreasing slope. Thus, it is concave.  \autoref{lemma:tildergeqigeqrgeqi} shows that the function $\tilde{R}_{\geq i}$ is always larger that $R_{\geq i}$. Note that the function $\tilde{R}_{\geq i}$  is piecewise linear and changes form at points $0, n, n+1, n+2, \cdots, n+i -1, n+i , 3n+i$, and $5n$.  \autoref{lemma:verticesmatch} shows that at all these points, the value of $\tilde{R}_{\geq i}$ is that same as that of $R_{\geq i}$. This finishes the proof.
\end{proof}

\subsection{Cleaning} \label{app:cleaning}

We start by fixing notation that we employ in this subsection. The letters $R$, $\tilde{R}$, $q$,  and $F$ are reserved for the revenue curves and the type distribution of our FedEx instance. These are described in \autoref{eq:rgeqi}, \autoref{eq:tildergeqi}, and \autoref{eq:typedis} respectively ($q$ is assumed to be uniform and is omitted throughout). The letters $A$ and $B$ would denote the allocation curves of mechanisms, {\em i.e.}, the allocation curve on day $i$ would be denoted by $A_i$ or $B_i$. Thus, for $x \in [0,v_{\max}]$, a bidder that reports type $(x,i)$ gets the item with probability $A_i(x)$.  We assume without loss of generality that $A_i(0) = 0$ and $A_i(v_{\max}) = 1$. In line with the notation used for probability distributions, we reserve the lower case $a_i$ for the `allocation density', {\em i.e.}, $a_i(x) = A_i(x) - A_i(x-1)$. With this notation, the menu complexity of a mechanism $M_A$ is the number of $(x,i)$ such that $a_i(x) \neq 0$. Unless specified otherwise, all statements about clean mechanisms pertain our particular instance of the FedEx problem defined in the previous section. 

In the FedEx problem, menus on day $i$ are constrained by the menu offered on day $i-1$. These constraints are represented by the downwards IC constraints in \autoref{eq:LP1}. In this sense, some menus on day $i-1$ are strictly more constraining than others. For example, a price of $p_1$ is offered on day $i-1$ is strictly more constraining than a price of $p_2 > p_1$.  This is because the utility $\pi(v,i-1)v - p(v,i-1)$ is strictly higher. Similarly, a distribution over two prices of $p_1$ and $p_2$ with probabilities $q$ and $1-q$ respectively is strictly more constraining than offering the single price $p= qp_1 + (1-q)p_2$.  This reasoning also applies in case the price $p$ has a small mass in a larger distribution. In the case a feasible less constraining menu generates a higher revenue on a particular day than a more constraining one, the latter can be changed to the former without violating any of the feasibility constraints. This change will only increase the revenue generated. 

In the problem instance we consider, the revenue curve $R_{i}$ for a given day $i$ is increasing and touches the upper concave envelope at all points in $\mathbb{Z} \cap [0,n+i]$. Thus, if a price $p$ in this range is offered with probability $f(p)$ on day $i-1$, the least constraining and revenue maximizing option on day $i$ is to offer the same price with the same probability. We define `clean' mechanisms to be mechanisms that satisfy this property. Formally,

\begin{definition}[Clean mechanisms] \label{def:cleaning} Let $a_i$ be the allocation density for day $i$ of a mechanism $M$. The mechanism $M$ is said to be clean if for all days $ i \geq 1 $ and any point $0\leq x \leq n+i$, we have
\[a_{i+1}(x) \geq a_{i}(x).\]
\end{definition}

The reason we do not require equality in the expression above is that the (allocation) mass from points higher than $n+i$ may be moved to lower points causing an increase in the value of $a_i(\cdot)$ there.  We prove in \autoref{thm:cleaning} that an arbitrary mechanism can be converted to a clean mechanism that is `intimately' connected to the original mechanism. %The proof of this theorem can be found in \autoref{app:cleaning}. 
We state a direct corollary here that is obtained by setting $k=n$ and $x_i =n+i$ for all $i$  in the statement of the theorem.

\begin{corollary}[Cleaning] \label{cor:cleaning} Consider any mechanism $M_A$ for the FedEx instance described in \autoref{sec:approxinstance}.  There exists a mechanism $M_B$ such that:
\begin{itemize}
\item $M_B$ mimics $M_A$ on day $1$.
\item $M_B$ is clean.
\item For all days $i$, we have $B_i \succeq A_i$.
\item For all days $i$, $\sum_{j=0}^{v_{\max}} R_i(j)b_i(j) \geq \sum_{j=0}^{v_{\max}} R_i(j)a_i(j)$.
\end{itemize}

Here, $\succeq$ denotes stochastic dominance of the second-order. 
\end{corollary}

\subsection{The statement and proof of \autoref{thm:cleaning} \protect \footnote{Auxiliary results that can be skipped without loss of comprehension.}}

Consider any mechanism $M$. Let $A_i(v) : \{0,1,\cdots, V\} \to [0,1]$ denote the allocation function of mechanism $M$ on day $i$. We know from the IC constraints that $A_i(v)$ is monotone non-decreasing for all $i$. Without loss of generality, assume that $A_i(V) = 1$. We interpret $A_i$ as the cumulative distribution function of some random variable. The leftwards IC constraints say that for all $i > 1, v \leq V$, we have,

\begin{equation} \label{eq:leftic}
\sum_{j = 0}^v A_i(j)  \geq \sum_{j=0}^v A_{i-1}(j).
\end{equation}

We define stochastic dominance and state a well-known result 

\begin{definition} \label{def:sosd} Consider two discrete random variables $X$ and $Y$ supported on the non-negative integers.  ($F$, $G$ are the cdfs of the random variables). We say $X$ has (second-order) stochastic dominance over $Y$ if and only if for all $x \in \mathbb{Z}$,

\[\sum_{j =0}^x F(j) \leq \sum_{j=0}^x G(j).\]

We denote this by $X \succeq Y$.
\end{definition}

We have the following well-known result on necessary and sufficient conditions for second-order stochastic dominance:

\begin{theorem} \label{thm:sosd} Let $X_A$ and $X_B$ be two random variables with distributions $A$ and $B$ respectively. The following statements are equivalent:

\begin{itemize}
\item $X_A \succeq X_B $.
\item There exist random variables $Y$ and $Z$ such that $X_B \overset{d}{=} X_A + Y+Z$, with $Y$ always at most $0$ and $Z$ such that $\mathbb{E}[Z \mid X_A+Y] = 0$.
\item For any concave and increasing function $f$, it holds that $\sum_{j=0}^Vf(j)a(j) \geq \sum_{j=0}^V f(j)b(j)$.
\end{itemize}

Here, $\overset{d}{=}$ denotes equivalence in distribution.
\end{theorem}

We proceed to prove our main technical lemma.

\begin{lemma} \label{lemma:sandwich} If any two random variables $A$ and $C$ defined on $[0,V]\cap \mathbb{Z}$ satisfy $A \succeq C$, then for any $x$, there exists a $B$ such that $A \succeq B \succeq C$ and for all $v \leq x$, we have $\Pr(B = v) \geq \Pr(A=v)$.

Furthermore, if $f$ is a function that is increasing on $[0,x]$ such that $f(j) = \tilde{f}(j)$ for all $0 \leq j \leq x$, then $\sum_{j=0}^V f(j)\Pr(B = j) \geq \sum_{j=0}^V f(j)\Pr(C= j)$.
\end{lemma}
\begin{proof} 

\autoref{thm:sosd} promises random variables $Y$ and $Z$ such that $C \overset{d}{=} A + Y + Z$, where $Y \leq 0$ and  $\mathbb{E}[Z \mid A+Y] = 0$. Our goal is to construct a random variable $B$ such that $A \succeq B \succeq C$. We wish to establish both the stochastic dominance results via \autoref{thm:sosd}. To this end, we first construct random variables $Y_1$ and $Z_1$  and use these to define $B$. Let

 \begin{equation} \label{eq:y1} 
 Y_1 = \begin{cases} 0, & \text{if } A < x\\
 \max(Y, x - A), & \text{if } A \geq x.\\
 \end{cases}
 \end{equation}
 
 Note that $Y_1 \leq 0$. Next, define $Z_1$

 \begin{equation} \label{eq:z1} 
 Z_1 = \begin{cases} Z, & \text{if }  x - A < Y_1\\
 0, & \text{if } x - A \geq Y_1.\\
 \end{cases}
 \end{equation}
 
 We prove that $\mathbb{E}[Z_1 \mid A + Y_1] = 0$ in \autoref{claim:validz1} paving the way for \autoref{thm:sosd}. For that, we need another result.
  
 \begin{claim}\label{claim:eventssame} For any $k > x$, the event $A + Y_1 = k$ happens if and only if $A + Y = k$.
 \end{claim}
 \begin{proof} First, assume $A + Y_1 = k  > x$. This means that $A \geq A + Y_1 = k  > x$ and hence, by \autoref{eq:y1}, $Y_1 = \max(Y, x - A)$. However, $Y_1 > x - A$ implying that $Y_1 = Y$.
 
 Second, assume that $A + Y = k$. Again, we have $A \geq A + Y = k  > x$ and hence, by \autoref{eq:y1}, $Y_1 = \max(Y, x - A) = Y$.
 \end{proof}
  
 \begin{claim} \label{claim:validz1} $\mathbb{E}[Z_1 \mid A + Y_1] = 0.$
 \end{claim}
 \begin{proof}  Fix $A + Y_1 = k$ where $k$ is arbitrary. If $k \leq x$, then $Z_1 = 0$ and the claim follows.
 
 If, on the other hand, $ k  > x$, then \autoref{claim:eventssame} applies. Thus, by \autoref{eq:z1}, we have
  
 \[\mathbb{E}[Z_1 \mid A + Y_1 = k] = \mathbb{E}[Z \mid A + Y = k] = 0.\]
 
 \end{proof}

Define $B = A + Y_1 + Z_1$.  Then, \autoref{claim:validz1} means that  $A  \succeq B $. Now, we wish to prove that $B \succeq C $. Our strategy is to again employ \autoref{thm:sosd}. For this, we define $Y_2$ and $Z_2$.

 \begin{equation} \label{eq:y2} 
 Y_2 = Y - Y_1.
 \end{equation}
 
 Note that

 \begin{claim} \label{claim:validy2} $Y_2 \leq 0$.
 \end{claim}
 \begin{proof}  We prove that $Y \leq Y_1$. If $A < x$, the $Y_1 = 0$ and this is trivial. Otherwise, $Y_1 =  \max(Y, x - A) \geq Y$.
  \end{proof}
 
 We next define $Z_2$ as

 \begin{equation} \label{eq:z2} 
 Z_2 = Z - Z_1.
 \end{equation}

 \begin{claim} \label{claim:validz2} $\mathbb{E}[Z_2 \mid B + Y_2] = 0$.
 \end{claim}
 \begin{proof}  Fix $B + Y_2 = k$ where $k$ is arbitrary. Since $B = A + Y_1 + Z_1$, this is equivalent to saying that $A + Y + Z_1 = k$.
 
 We prove that $\mathbb{E}[Z_2 \mid B + Y_2] = 0$ by arguing $\mathbb{E}[Z_2 \mid B + Y_2, x - A < Y_1] = 0$ and $\mathbb{E}[Z_2 \mid B + Y_2, x - A \geq Y_1] = 0$. We first deal with the former. If $x - A < Y_1$, then $Z_1 = Z$ (\autoref{eq:z1}) and hence $Z_2 = 0$ and we are done. 
 
 Now, suppose $x - A \geq Y_1$. In this case, $Z_1 = 0$ (\autoref{eq:z1}) and hence $B+Y_2 = k$ if and only if $A+Y = k$.
 %$Z_2 = Z$ and $A + Y = k$ (\autoref{eq:z1}). \rrsnote{ Note that this is if and only if.} Thus, 
 
 \[\mathbb{E}[Z_2 \mid B + Y_2 = k , x - A \geq Y_1] = \mathbb{E}[Z \mid A + Y = k , x - A \geq Y_1]. \]
 
 Note that $ x - A \geq Y_1$ is the same as $ x \geq A + Y_1$ which happens if and only if $ x \geq A + Y = k$ (\autoref{claim:eventssame}). Thus, 
 
 \[\mathbb{E}[Z \mid A + Y = k , x - A \geq Y_1]  = \mathbb{E}[Z \mid A + Y = k , x \geq k] = 0.\]

 \end{proof}

Since $B+Y_2 + Z_2 = A+ Y_1 + Z_1 + Y_2 +Z_2 = A+Y+Z = C$ by \autoref{eq:y2} and \autoref{eq:z2}, we have that $B \succeq C$ by \autoref{claim:validy2} and \autoref{claim:validz2}.

Consider the event $A = v$ for $v < x$. Since $v \leq x$, both $Y_1$ and $Z_1$ are $0$. Thus, $B = v$.  Hence, $\Pr(B=v) \geq \Pr(A=v)$.

Finally, consider a function $f$ such that $f(j) = \tilde{f}(j)$ for all $0 \leq j \leq x$. We wish to prove that $\sum_{j=0}^V f(j)\Pr(C= j)  \leq \sum_{j=0}^V f(j)\Pr(B= j) $. We break the proof into $2$ cases. Each one of the following claims handles one such case.

\begin{claim} Let $E_1$ be the event $ A+Y > x$. We have $\sum_{j=0}^V f(j)\Pr(C= j \mid E_1)  \leq \sum_{j=0}^V f(j)\Pr(B= j \mid E_1)$
\end{claim}
\begin{proof} By \autoref{claim:eventssame}, $E_1$ can equivalently be described as $A+Y_1 > x$. We use these two forms interchangeably in this proof. If $E_1$ occurs, $Y_1 = Y$ and $Z_1 = Z$ (\autoref{eq:y1} and \autoref{eq:z1}). Thus, $B=C$ and the claim follows.
\end{proof}

\begin{claim} Let $E_2$ be the event $A+Y \leq x$. We have $\sum_{j=0}^V f(j)\Pr(C= j \mid E_2)  \leq \sum_{j=0}^V f(j)\Pr(B= j \mid E_2)$
\end{claim}
\begin{proof} By \autoref{claim:eventssame}, $E_2$ can equivalently be described as $ A+ Y_1 \leq x$. We use these two forms interchangeably in this proof. If $E_2$ occurs,  $Z_1 = 0$ (\autoref{eq:z1}). Thus, $B=  A+ Y_1 + Z_1 = A+Y_1$ implying $x \geq B \geq A + Y$. Thus, for all $j_1 \leq x$, we have  

\[\sum_{j=0}^V f(j)\Pr(B= j \mid A+Y = j_1) = \sum_{j=j_1}^x f(j)\Pr(B= j \mid A+Y = j_1) \geq f(j_1).\] 

We have 

\begin{align*}
\sum_{j=0}^V f(j)\Pr(C= j \mid E_2) &\leq  \sum_{j=0}^V \tilde{f}(j)\Pr(C= j \mid E_2)\\ 
&\leq  \sum_{j_1 = 0}^x \Pr(A+Y = j_1 \mid E_2) \sum_{j=0}^V \tilde{f}(j)\Pr(C= j \mid A+Y = j_1)\\  
&\leq  \sum_{j_1 = 0}^x \Pr(A+Y = j_1 \mid E_2) \tilde{f}\left( \sum_{j=0}^Vj\Pr(C= j \mid A+Y = j_1)\right)\\ 
&=  \sum_{j_1 = 0}^x \Pr(A+Y = j_1 \mid E_2) \tilde{f}\left( j_1\right)\\  
&\leq  \sum_{j_1 = 0}^x \Pr(A+Y = j_1 \mid E_2) \sum_{j=0}^V\tilde{f}\left( j\right) \Pr(B = j \mid A+Y = j_1)\\ 
\end{align*}

If $A+Y = j_1 \leq x$, we have $B \in [j_1, x]$ and hence, 
\begin{align*}
\sum_{j=0}^V f(j)\Pr(C= j \mid E_2) &\leq  \sum_{j_1 = 0}^x \Pr(A+Y = j_1 \mid E_2) \sum_{j=0}^V\tilde{f}\left( j\right) \Pr(B = j \mid A+Y = j_1)\\ 
&=  \sum_{j_1 = 0}^x \Pr(A+Y = j_1 \mid E_2) \sum_{j=0}^Vf\left( j\right) \Pr(B = j \mid A+Y = j_1)\\   
&=  \sum_{j=0}^Vf\left( j\right) \Pr(B = j \mid E_2)\\  
\end{align*}

\end{proof}

\end{proof}

We are now ready to prove our main ``cleaning'' result:
\begin{theorem}[Cleaning] \label{thm:cleaning} Consider any mechanism $M_A$ for any FedEx instance with type space $[0,V] \times \{1,2,\cdots n\}$. Let $\{x_i\}_{i=1}^{n-1}$ be a sequence of numbers in $[0,V]$.  For any $k \geq 1$, there exists a mechanism $M_B$ such that:
\begin{itemize}
\item $M_B$ mimics $M_A$ on day $1$.
\item For all $1 \leq  i \leq k-1$ and $v \leq x_i$,we have $b_{i+1}(v) \geq b_i(v)$.
\item For all $1 \leq i \leq k$, we have $B_i \succeq A_{i}$ and for all $i > k$ we have $B_i = A_i$.
\item For all $1 < i \leq k$, if the revenue curve $R_i$ on day $i$ of the bidder is increasing on $[0,x_{i-1}]$ such that $R_i(j) = \tilde{R_i}(j)$ for all $0 \leq j \leq x_{i-1}$, then $\sum_{j=0}^V R_i(j)b_i(j) \geq \sum_{j=0}^V R_i(j)a_i(j)$.
\end{itemize}
\end{theorem} 
If it were not for the last two bullets, \autoref{thm:cleaning} would be trivial. We could have set $M_B$ to be the optimal mechanism given the allocation of $M_A$  on day $1$. That the optimal mechanism is clean is an easy consequence of Fiat et al's algorithm~\cite{Fiat:2016}. Intuitively, these bullets say that the curves $A_i$ can be obtained from $B_i$ by `splitting' and `lowering' prices appropriately and that these actions do not decrease the revenue generated.  This structure ensures that we can convert results for the mechanism $M_B$ into results for $M_A$.

This is exactly how we proceed. We take an arbitrary auction and clean it. We prove that a clean auction that generates high revenue must have a high menu complexity. We then translate this into a result for the original auction.

\begin{proof}[Proof of \autoref{thm:cleaning}]
Proof by induction on $k$. If $k=1$, we set $M_B = M_A$ and observe that all the conditions are satisfied. We assume the result holds for $k-1$ and prove it for $k>1$. Let $M_C$ be the mechanism promised by the induction hypothesis. 
Since $M_C$ is feasible, we have 
\[C_{k-1} \succeq C_k.\]
Applying \autoref{lemma:sandwich} on $C_{k-1}, C_{k} = A_k$ and $x_{k-1}$ , we get $B_{k}$ such that 
\begin{equation} \label{eq:sandwich}
C_{k-1} \succeq B_k \succeq A_k ,
\end{equation}
and for all $v \leq x_{k-1}$, we have $b_k(v) \geq c_{k-1}(v)$. Furthermore, if $f$ is a function that is increasing on $[0,x_{k-1}]$ such that $f(j) = \tilde{f}(j)$ for all $0 \leq j \leq x_{k-1}$, then $\sum_{j=0}^V f(j)b_k(j) \geq \sum_{j=0}^V f(j)a_k(j)$. Define the mechanism $M_B$ to be $M_C$ with the allocation $C_k$ on day $k$ replaced by $B_k$. \autoref{eq:sandwich} implies that $M_B$ is feasible.

Observe that $M_B$ satisfies all requirements.

\end{proof}

\subsection{Analysis of clean auctions}

If $f_i$ is the marginal density of bidder types on day $i$, then the revenue generated by the allocation density $a_i$  is equal to $\sum_{x=0}^{v_{\max}} f_i(x)p_i(x)$ where $p_i(x)$ is the payment made by the bidder that reports type $(x,i)$.  Due to Myerson's payment identity~\cite{Myerson81}, we know that $p_i(x) = \sum_{j=0}^x  j a_i(j)$. Using this expression, the revenue generated by the allocation density $a_i$ on day $i$ is equal to 

\begin{align*}
\sum_{x=0}^{v_{\max}} f_i(x)p_i(x) &= \sum_{x=0}^{v_{\max}}  \sum_{j=0}^x   j a_i(j) f_i(x) \\
&= \sum_{j=0}^{v_{\max}}  j a_i(j) (1-F(j-1)) = \sum_{j=0}^{v_{\max}} a_i(j) R_i(j). \\
\end{align*}

We will use this expression frequently. \autoref{lem:geqi} says that given the allocation density $a_i$, the revenue generated by the optimal mechanism on days $i$ through $n$ is $\sum_{j=0}^{v_{\max}} a_i(j) R_{\geq i}(j) $. Of this revenue, an amount equal to $\sum_{j=0}^{v_{\max}} a_i(j) R_i(j) $ is generated on day $i$ and the remainder $\sum_{j=0}^{v_{\max}} a_i(j) R_{\geq i}(j) - \sum_{j=0}^{v_{\max}} a_i(j) R_i(j) $ is generated on days $i+1$ and onwards. By \autoref{def:revcurves}, this is equal to $\sum_{j=0}^{v_{\max}} a_i(j) \tilde{R}_{\geq i+1}(\min(j,3n+i+1)) $.

\subsubsection{Analyzing clean auctions}

We described the optimal auction for our instance. It has the $i-1$ menu options $n+2, n+3, \cdots , n+i-1$ and $3n+i-1$ on day $i-1$. On day $i$, the menu option $3n+i-1$ is split into $n+i$ and $3n+i$ while all the others are carried over. The first lemma we prove tries to formalize that this split is necessary. We intend to use this lemma in our analysis of clean auctions. By definition, a clean auction carries all prices in the range $[0,n+i-1]$ on day $i-1$ over to day $i$. We prove that if a price between $n+i$ and $3n+i-1/2$ is split without creating a new menu option of $n+i$, a significant revenue loss is incurred.  This loss is the difference between what is actually obtained and what what could have been obtained.

If a price of $x < 3n+i$ is set on day $i-1$, then the maximum revenue that can be generated day $i$ onwards is  $\tilde{R}_{\geq i}(x)$. The revenue generated by an allocation $a_i$ is at most $\sum_{j=0}^{v_{\max}} a_i(j) R_{\geq i}(j)$. We prove that if an allocation is feasible (this translates to $\sum_{j=0}^{v_{\max}} j a_i(j) \leq x$) but $a_i(n+i) = 0$, then the difference between the two terms above is large.

\begin{lemma} \label{lemma:pointmass} Fix any $1\leq i \leq n$ and $n+i \leq x \leq 3n+i-1/2$. For any allocation density $a_i$, if $a_i(n+i) = 0$ and $ \sum_{j=0}^{v_{\max}} j a_i(j) \leq x$, it holds that
\[\tilde{R}_{\geq i}(x) - \sum_{j=0}^{v_{\max}} a_i(j) R_{\geq i}(j) \geq \frac{n+1-i}{10(2n+1)}.\]
\end{lemma}
\begin{proof} 

Consider the line joining the points $(n+i-1, \tilde{R}_{\geq i}(n+i-1))$ and $(3n+i, \tilde{R}_{\geq i}(3n+i))$. Any point on this line satisfies $y = \ell(x)$ for a linear function $\ell$. Define the function $S(x) = \min(\ell(x), \tilde{R}_{\geq i}(x))$. The function $S$ is the minimum of two concave functions and hence is concave. Further, $S(x) \geq R_{\geq i}(x)$ on all points $x \in \{0,1,2, \cdots, 5n\} \setminus \{n+i\}$.\footnote{This can be verified by the interested reader. We omit the calculations.} Using these facts and Jensen's inequality,

\[\sum_{j=0}^{v_{\max}} a_i(j) R_{\geq i}(j) \leq \sum_{j=0}^{v_{\max}} a_i(j) S(j)\leq  S\left(\sum_{j=0}^{v_{\max}} a_i(j) j\right).\]

We observe that the function $S$ is increasing below $3n+i$. Thus, 

\[\sum_{j=0}^{v_{\max}} a_i(j) R_{\geq i}(j) \leq S\left(\sum_{j=0}^{v_{\max}} a_i(j) j\right) \leq S(x) .\]

By direct calculation, we have that

\begin{align*}
\sum_{j=0}^{v_{\max}} a_i(j) R_{\geq i}(j) &\leq S(x) \leq \ell(x)\\
&= \frac{3n+i-x}{2n+1}\tilde{R}_{\geq i}(n+i-1) + \frac{x+1-n-i}{2n+1}\tilde{R}_{\geq i}(3n+i) \\
&= \frac{3n+i-x}{2n+1}(n+1-i)S_{i-1} + \frac{x+1-n-i}{2n+1}(n+1-i)(n+S_i) &\text{(\autoref{eq:tildergeqi})}\\
%&= \frac{n+1-i}{2n+1}\left((3n+i-x)S_{i-1} + (x+1-n-i)(n+S_i) \right)\\
%&= \frac{n+1-i}{2n+1}\left((3n+i)S_{i-1} - (n+i-1)(n+S_i) + x(n+S_i - S_{i-1}) \right)\\
%&= \frac{n+1-i}{2n+1}\left((2n+1)S_{i} - (n+i-1)n + x\left(n+\frac{1}{\lambda^{i-1}}\right) - (3n+i)\frac{1}{\lambda^{i-1}} \right)\\
%&= \frac{n+1-i}{2n+1}\left((2n+1)S_{i} + (x-n-i)n  +n - (3n+i-x)\frac{1}{\lambda^{i-1}} \right)\\
&=  (n+1-i)\left(C(x-n-i)+S_i  + \frac{3n+i-x}{2n+1}\left(\frac{1}{2} - \frac{1}{\lambda^{i-1}} \right)\right).\\
\end{align*}

We also have that

\begin{align*}
\tilde{R}_{\geq i}(x) &=  (n+1-i)\left(C(x-n-i)+S_i \right). &\text{(\autoref{eq:tildergeqi})}\\
\end{align*}

Combining, 

\begin{align*}
\sum_{j=0}^{v_{\max}} a_i(j) R_{\geq i}(j)   &\leq \tilde{R}_{\geq i}(x) - \frac{n+1-i}{2n+1}(3n+i-x)\left(\frac{1}{\lambda^{i-1}} -\frac{1}{2} \right)\\
&\leq \tilde{R}_{\geq i}(x) - \frac{n+1-i}{2(2n+1)}\left(\frac{1}{\lambda^{i-1}} -\frac{1}{2} \right)\\
&< \tilde{R}_{\geq i}(x) - \frac{n+1-i}{10(2n+1)}. & \text{(Claim~\ref{claim:lambdapowers})}\\
\end{align*}

\end{proof}

A corollary of the above result takes care of the case where $a_i(n+i) \neq 0$ but is small.

\begin{corollary} \label{cor:pointmass} Fix any $1\leq i \leq n$ and $n+i \leq x \leq 3n+i-3/4$. For any allocation density $a_i$, if $a_i(n+i) < \frac{1}{100n}$ and $ \sum_{j=0}^{v_{\max}} j a_i(j) \leq x$, it holds that
\[\tilde{R}_{\geq i}(x) - \sum_{j=0}^{v_{\max}} a_i(j) R_{\geq i}(j) \geq \frac{n+1-i}{20(2n+1)}.\]
\end{corollary}
\begin{proof} 

Construct the allocation  \[a'_i(x) = \begin{cases} 0, &\text{if } x = n+i\\\frac{a_i(x)}{1-a_i(n+i)}, &\text{if } x \neq n+i.\end{cases}\]

Note that:

\begin{equation}\label{eq:changeinexpec}
(1-a_i(n+i))\left(\sum_{j=0}^{v_{\max}} j a'_i(j) - (n+i)\right) = \sum_{j=0}^{v_{\max}} j a_i(j) - (n+i). \\
\end{equation}
Set $y = \max\left( \sum_{j=0}^{v_{\max}} j a'_i(j), n+i\right) \geq n+i$. Note that $y \leq 3n+i-1/2$ because:

\begin{align*}
\sum_{j=0}^{v_{\max}} j a'_i(j) &= \frac{1}{1-a_i(n+i)}\left(\sum_{j=0}^{v_{\max}} j a_i(j) - (n+i)\right) + (n+i)\\
&\leq \frac{100n}{100n-1}\left(2n - 3/4\right) + (n+i) \leq 3n+i - 1/2.\\
\end{align*}

Thus, \autoref{lemma:pointmass} holds for $y, a'_i$ and we have
\begin{align*}
\sum_{j=0}^{v_{\max}} a_i(j) R_{\geq i}(j) &- R_{\geq i}(n+i) \\
&=  \left(1-a_i(n+i)\right)\left(\sum_{j=0}^{v_{\max}} a'_i(j) R_{\geq i}(j)- R_{\geq i}(n+i)\right)\\
&\leq \left(1-a_i(n+i)\right)\left(\tilde{R}_{\geq i}(y) - R_{\geq i}(n+i) - \frac{n+1-i}{10(2n+1)}\right) &\text{(Lemma~\ref{lemma:pointmass})}  \\
&= \left(1-a_i(n+i)\right)\left((n+1-i)C\left(y  - (n+i)\right) - \frac{n+1-i}{10(2n+1)}\right)  &\text{(\autoref{eq:tildergeqi})}   \\
&\leq  (n+1-i)C\left(x - (n+i)\right) - \left(1-a_i(n+i)\right) \frac{n+1-i}{10(2n+1)} &\text{(\autoref{eq:changeinexpec})}    \\
&=  \tilde{R}_{\geq i}(x)  - R_{\geq i}(n+i) - \left(1-a_i(n+i)\right) \frac{n+1-i}{10(2n+1)}. &\text{(\autoref{eq:tildergeqi})}    \\
\end{align*}

Adding $R_{\geq i}(n+i) $ to both sides, we get the result. 
\end{proof}

\autoref{cor:pointmass} is our final result about the prices that split. These prices (in our instance) are in the range $[n+i, 3n+i]$ on day $i$.  We next state results concerning prices outside this range. First, we deal with points $x > 3n+i$ such that $a_i(x) > 0$. The next pair of lemmas we prove upper bound the `allocation mass' on this tail.

\begin{lemma} \label{lemma:gapaftermax} For any $1\leq i \leq n$ and $x>3n+i$, $R_{\geq i}(3n+i) -  R_{\geq i}(x) \geq n$. 
\end{lemma}
\begin{proof} Proof by calculation. 
\begin{align*}
R_{\geq i}(3n+i) - R_{\geq i}(x) &\geq \beta_i(3n+i) + (n-i)(C(2n-1) + S_i) - (n-i)(2nC+ S_i) &\text{(\autoref{eq:rgeqi})}\\
&= \beta_i(3n+i) - (n-i)C\\
& > \frac{3n+i}{2} - \frac{n-i}{2} &\text{(Claim~\ref{claim:betarange})}\\
&= n+i > n.\\
\end{align*}
\end{proof}

The next lemma shows that any approximately optimal mechanism for our instance must allocate very little mass to the tail end of the distribution.

\begin{lemma} \label{lemma:nomassaftermax} Consider any mechanism $M$ for the FedEx instance described in Section~\ref{sec:approxinstance}. If $M$ generates at least a $1-\frac{1}{1000n^2}$ fraction of the optimal revenue $\mathsf{OPT}$, then on all days $1 \leq i \leq n$, the allocation curve $A_i$ of the mechanism $M$ satisfies,
\[A_i(3n+i) \geq 1-\frac{1}{200n}.\]  
\end{lemma}
\begin{proof} The statement can be verified for $i =1$ and so we assume $i > 1$. Consider the allocation curve $A_{i-1}$ on the $(i-1)^{\text{th}}$ deadline. Given $A_{i-1}$, the optimal revenue that can be generated on days $i$ through $n$ is $\sum_{j=0}^{v_{\max}} a_{i-1}(j) \tilde{R}_{\geq i}(\min(j,3n+i)) $. Since $A_{i-1} \succeq A_i$, we know by the third bullet in \autoref{thm:sosd} that:

\[\sum_{j=0}^{v_{\max}} a_{i-1}(j) \tilde{R}_{\geq i}(\min(j,3n+i)) \geq \sum_{j=0}^{v_{\max}} a_{i}(j) \tilde{R}_{\geq i}(\min(j,3n+i)).\]

The actual revenue generated on days $i$ through $n$ is at most $\sum_{j=0}^{v_{\max}} a_i(j) R_{\geq i}(j) $. Thus, allocating according to $a_i$ on day $i$ incurs a revenue loss of at least $\sum_{j=0}^{v_{\max}} a_{i}(j) \tilde{R}_{\geq i}(\min(j,3n+i)) - \sum_{j=0}^{v_{\max}} a_i(j) R_{\geq i}(j) $ on days $i$ through $n$. This is at least:

\begin{align*}
\sum_{j=0}^{v_{\max}} a_{i}(j) \tilde{R}_{\geq i}(\min(j,3n+i)) &- \sum_{j=0}^{v_{\max}} a_i(j) R_{\geq i}(j) \\
 &\geq \sum_{j=3n+i+1}^{v_{\max}} a_{i}(j) \left(\tilde{R}_{\geq i}(3n+i) - R_{\geq i}(j)\right)  & \text{(As $\tilde{R}_{\geq i}(j) \geq R_{\geq i}(j)$)}\\
&\geq n \sum_{3n+i+1}^{v_{\max}}  a_{i}(x)dx = n\left(1-A_i(3n+i)\right). & \text{(\autoref{lemma:gapaftermax})}\\
\end{align*}

For a $1-\frac{1}{1000n^2}$ approximate auction, the loss in revenue is upper bounded by $\frac{1}{200}$. Thus, $n\left(1-A_i(3n+i)\right) \leq \frac{1}{200}$ implying $A_i(3n+i) \geq 1-\frac{1}{200n}$.

\end{proof}

Finally, we prove some technical results for allocating in the range $[0, n+i]$.

\begin{lemma}\label{lemma:lowmassfirstsegment} For any clean mechanism $M$ that generates revenue at least $\frac{99}{100}$ times $\mathsf{OPT}$, 
\[A_{n/3}(4n/3) \leq \frac{1}{2}.\]
\end{lemma}
\begin{proof} The Myerson optimal revenue for day $i$ is $\max_x R_i(x) = \beta_i(3n+i)$ by \autoref{eq:ri}. We upper bound the revenue generated by $M$ on days $1$ through $n/3 - 1$ by $\sum_{i=1}^{n/3 - 1} \beta_i(3n+i)$. We have 
\begin{align*} 
\sum_{i=1}^{n/3 - 1} \beta_i(3n+i) &= \sum_{i=1}^{n/3}\frac{3n-i}{2} - \frac{n-i}{\lambda^{i}} + S_i &\text{(\autoref{eq:beta})}\\
&\leq \sum_{i=1}^{n/3 - 1}\frac{3n-i}{2} - \frac{3(n-i)}{4} + n+i&\text{(\autoref{claim:lambdapowers})}\\
&\leq \sum_{i=1}^{n/3 - 1}\frac{7n+5i}{4} < \frac{3n^2}{4} - \frac{9n}{4}.\\
\end{align*}

The revenue generated on day $n/3$ onwards is upper bounded by $\sum_{j=0}^{v_{\max}} a_{n/3}(j) R_{\geq n/3}(j) $. Suppose for the sake of contradiction that $A_{n/3}(4n/3) > \frac{1}{2}$. If this is the case, then $\sum_{j=0}^{v_{\max}} a_{n/3}(j) R_{\geq n/3}(j)$ is upper-bounded by 

\begin{align*}
\sum_{j=0}^{v_{\max}} a_{n/3}(j) R_{\geq n/3}(j) &\leq A_{n/3}(4n/3) R_{\geq n/3}(4n/3) + (1-A_{n/3}(4n/3)) R_{\geq n/3}(10n/3) \\
&\leq R_{\geq n/3}(10n/3) - A_{n/3}(4n/3)\left( R_{\geq n/3}(10n/3)  - R_{\geq n/3}(4n/3)\right)  \\
&\leq \frac{11n^2}{9} + \frac{11n}{6}.  \\
\end{align*}

The sum of these two contributions is at most $\frac{3n^2}{4} + \frac{11n^2}{9} = \frac{71n^2}{36} \leq \frac{71}{72}\mathsf{OPT}$, a contradiction.
\end{proof}

\begin{lemma}\label{lemma:cuttail} Let $a_i$ and $a_{i+1}$ be allocations that satisfy $ \sum_{j=0}^x A_{i+1}(j) \geq  \sum_{j=0}^x  A_i(j)$ for all $x$. Further, let $v_i, v_{i+1}$ be such that $A_i(v_i) = A_{i+1}(v_{i+1}) = \alpha$. Then, 
\[\sum_{j=0}^{v_{i+1}} ja_{i+1}(j) \leq \sum_{j=0}^{v_i} ja_{i}(j). \]  
\end{lemma}
\begin{proof}  
First suppose that $v_i < v_{i+1}$. We have:
\begin{align*}
\sum_{j=0}^{v_{i+1}} ja_{i+1}(j) - \sum_{j=0}^{v_i} ja_{i}(j) &= \alpha (v_{i+1}- v_i) - \sum_{j=0}^{v_{i+1}} A_{i+1}(j) + \sum_{j=0}^{v_{i}} A_i(j)\\
 &\leq \alpha (v_{i+1}- v_i) - \sum_{j=v_i+1}^{v_{i+1}} A_{i}(j)\\
 &\leq \alpha (v_{i+1}- v_i) - \sum_{j=v_i+1}^{v_{i+1}} \alpha = 0.\\
\end{align*}
If $v_i \geq v_{i+1}$. We have:
\begin{align*}
\sum_{j=0}^{v_{i+1}} ja_{i+1}(j) - \sum_{j=0}^{v_i} ja_{i}(j) &= \alpha (v_{i+1}- v_i) - \sum_{j=0}^{v_{i+1}} A_{i+1}(j) + \sum_{j=0}^{v_{i}} A_i(j)\\
 &\leq \alpha (v_{i+1}- v_i) + \sum_{j=v_{i+1}+1}^{v_{i}} A_{i}(j)\\
 &\leq \alpha (v_{i+1}- v_i) + \sum_{j=v_{i+1}+1}^{v_{i}} \alpha = 0.\\
\end{align*}
\end{proof}

We are now ready to prove a menu complexity lower bound for clean auctions.

\begin{theorem} \label{thm:cleanauctions} Consider any clean mechanism $M$ for the FedEx instance described in \autoref{sec:approxinstance} that is $1-\frac{1}{1000n^2}$ approximate. For any day $i$ such that $n/4 \leq i < n/2$, the mechanism $M$ satisfies $a_i(j) \geq \frac{1}{300n}$ for all $j \in [n+2,n+n/4]$.
\end{theorem}

\begin{proof} Consider any day $i < n/4$ and the pair $a_{i}$ and $a_{i+1}$ of allocation densities of the mechanism $M$ on day $i$ and day $i+1$ respectively. Given $a_{i}$, the optimal revenue that can be generated on days $i+1$ through $n$ is $\sum_{j=0}^{v_{\max}} a_i(j) \tilde{R}_{\geq i+1}(\min(j,3n+i+1)) $. The actual revenue generated on days $i+1$ through $n$ is at most $\sum_{j=0}^{v_{\max}} a_{i+1}(j) R_{\geq i+1}(j) $. Thus, allocating according to $a_{i+1}$ on day $i+1$ incurs a revenue loss of at least $\sum_{j=0}^{v_{\max}} a_i(j) \tilde{R}_{\geq i+1}(\min(j,3n+i+1)) - \sum_{j=0}^{v_{\max}} a_{i+1}(j) R_{\geq i+1}(j) $ on days $i+1$ through $n$.

%This is at least:
%
%
%\begin{equation} \label{eq:loss1}
%\int_0^{v_{\max}} \tilde{R}_{\geq i+1}(x)a_{i}(x)dx - \int_0^{v_{\max}} R_{\geq i+1}(x)a_{i+1}(x)dx.
%\end{equation}

Let $v^*$ be such that $A_i(3n+i) = A_{i+1}(v^*)$. We define two allocations $b_i$ and $b_{i+1}$\footnote{In this definition we implicitly assume that $n+i \leq v^*$. This holds because $A_{i+1}(n+i) \leq  \frac{1}{2}$ and $A_{i+1}(v^*)  = A_i(3n+i) \geq 1-\frac{1}{200n}$ by \autoref{lemma:nomassaftermax}. Since the mechanism is clean and $i < n/4 < n/3$, any `allocation mass' on day $i+1$ on points $[0,n+i+1]$ is pushed forward to all days. Thus, $A_{i+1}(n+i) \leq A_{i+1}(n+i+1) \leq A_{n/3}(4n/3) \leq \frac{1}{2}$ by \autoref{lemma:lowmassfirstsegment}.}.
\[b_i(v) = \begin{cases} 0,& v \leq n+i\\Z_1 a_{i}(v), & n+i < v \leq 3n+i \\ 0, & 3n+i<v, \end{cases} \hspace{2cm} b_{i+1}(v) = \begin{cases} Z_2(a_{i+1}(v) - a_i(v)), & v \leq n+i\\Z_2 a_{i+1}(v), & n+i < v \leq v^*\\ 0, & v^* < v, \end{cases} \]
where $Z_1$, $Z_2$ are the normalization constants. Since the mechanism $M$ is clean, we have $a_{i+1}(v) \geq a_i(v)$ for all $v \leq n+i$ by \autoref{def:cleaning} and therefore $b_i$ and $b_{i+1}$ are valid allocation densities.

\begin{claim} $Z_1 = Z_2 \leq 3$.
\end{claim}
\begin{proof} We first prove that $Z_1 = Z_2$. This is because 
\[\sum_{j=0}^{v_{\max}} \frac{b_i(j)}{Z_1} = A_i(3n+i) - A_i(n+i) = A_{i+1}(v^*) - A_i(n+i) = \sum_{j=0}^{v_{\max}} \frac{b_{i+1}(j)}{Z_2}.\]

We now prove that $Z_1 < 3$. Since the mechanism is clean and $i < n/4 < n/3$, any `allocation mass' on points $[0,n+i]$ is pushed forward to all days. Thus, $A_i(n+i) \leq A_{n/3}(4n/3) \leq \frac{1}{2}$ by \autoref{lemma:lowmassfirstsegment}. Also, by \autoref{lemma:nomassaftermax}, $A_i(3n+i) \geq 1-\frac{1}{200n}$. Thus,

\[Z_1 = \frac{1}{A_i(3n+i) - A_i(n+i)} \leq \frac{1}{1/2-\frac{1}{200n}} < 3.\]
\end{proof}  

For the remainder of this proof, we set $Z$ denote the common value of $Z_1$ and $Z_2$. We are now ready to lower bound the loss in revenue caused by allocating according to $a_{i+1}$.  Since $\tilde{R}_{\geq i+1} = R_{\geq i+1}$  for all $j \leq n+i$, we get

\begin{equation} \label{eq:loss1}
\begin{aligned}
\sum_{j=0}^{v_{\max}} &a_i(j) \tilde{R}_{\geq i+1}(\min(j,3n+i+1)) - \sum_{j=0}^{v_{\max}} a_{i+1}(j) R_{\geq i+1}(j)\\
&\geq \frac{1}{Z}\left(\sum_{j=n+i+1}^{3n+i} b_i(j) \tilde{R}_{\geq i+1}(j) - \sum_{j=0}^{v^*} b_{i+1}(j) R_{\geq i+1}(j)\right)\\
&= \frac{1}{Z}\left(\tilde{R}_{\geq i+1}\left(\sum_{j=n+i+1}^{3n+i} j b_i(j) \right) - \sum_{j=0}^{v^*} b_{i+1}(j) R_{\geq i+1}(j)\right)&\text{($\tilde{R}_{\geq i+1}$ is linear in $[n+i+1,3n+i+1]$)}\\
\end{aligned}
\end{equation}
%
%\begin{equation} \label{eq:loss2}
%\begin{aligned}
%&\geq \frac{1}{Z}\left(\tilde{R}_{\geq i+1}\left(\sum_{j=n+i}^{3n+i} j b_i(j) \right) -  R_{\geq i+1}\left(y\right) + \frac{n-i}{20(2n+1)}\right), &\text{(\autoref{cor:pointmass})}\\
%\end{aligned}
%\end{equation}
%\rrsnote{Clean the last inequality}

We next plan to upper bound $ \sum_{j=0}^{v^*} b_{i+1}(j) R_{\geq i+1}(j)$ using \autoref{cor:pointmass}.   To this end, fix $y = \max\left(n+i+1, \sum_{j=n+i+1}^{3n+i} j b_i(j)\right)$. Note that $n+i+1 \leq y \leq 3n+i$.  We first  show that  $y \geq \sum_{j=0}^{3n+i} j b_i(j) \geq \sum_{j=0}^{v^*} j b_{i+1}(j)$. Adding $\sum_{j=0}^{n+i} Za_i(v)$ on both sides, we get that this is equivalent to showing that $\sum_{j=0}^{3n+i} Z j a_i(j) \geq \sum_{j=0}^{v^*} Z j a_{i+1}(j)$. This holds due to \autoref{lemma:cuttail}. Now suppose $b_{i+1}(n+i+1) < \frac{1}{100n}$. If this is the case, apply \autoref{cor:pointmass} to the last term in \autoref{eq:loss1} to get

\begin{align*}
\sum_{j=0}^{v_{\max}} &a_i(j) \tilde{R}_{\geq i+1}(\min(j,3n+i+1)) - \sum_{j=0}^{v_{\max}} a_{i+1}(j) R_{\geq i+1}(j)\\
&\geq \frac{1}{Z}\left(\tilde{R}_{\geq i+1}\left(\sum_{j=n+i}^{3n+i} j b_i(j) \right) -  R_{\geq i+1}\left(y\right) + \frac{n-i}{20(2n+1)}\right)\\
&\geq \frac{1}{3}\frac{n-i}{20(2n+1)} > \frac{1}{240}.\\
\end{align*}

where the last inequality is because $y \geq \sum_{j=0}^{3n+i} j b_i(j)$ and $\tilde{R}_{\geq i}$ is increasing below $3n+i$. 

Since $\frac{1}{240} > \frac{\mathsf{OPT}}{1000n^2}$, this is unaffordable. Thus, $b_{i+1}(n+i+1) \geq \frac{1}{100n}$ or $a_{i+1}(n+i+1) \geq \frac{1}{Z100n} > \frac{1}{300n}$. Since the mechanism is clean, this is pushed forward to all days implying the result.

\end{proof}

\subsubsection{Analyzing general auctions}

 \autoref{thm:cleanauctions} proves an $\Omega(n^2)$-menu complexity lower bound for clean auctions. In this section, we wish to extend this lower bound to general auctions. Our proof will rely heavily on \autoref{cor:cleaning}  which says that any auction can be viewed as the `muddled' version of a clean auction. More precisely, the allocation curves $A_i$ of any mechanism are stochastically dominated by clean curves $B_i$. We prove that removing a lot of menu options from $B_i$ to get $A_i$ while maintaining $B_i \succeq A_i$ results in a huge revenue loss day $i+1$ onwards. Since the revenue generated by $B_j$ for $j \leq i$ is at least that generated by $A_j$($4^{\text{th}}$ bullet in \autoref{cor:cleaning}), the huge loss of revenue on days $i+1$ onwards is tantamount to a huge loss overall.

 \begin{proof}[Proof of \autoref{thm:lowerbound}]
 Consider an arbitrary mechanism $M_A$ and let $A_i$ be the allocation curve of $M_A$ on day $i$. \autoref{cor:cleaning} says that there exist allocation curves $B_i$ of a clean mechanism $M_B$ such that $B_i \succeq A_i$ for all $i$. Let $i \in (n/4, n/2]$ be such that the number of menu options on day $i$ is at most $n/8$. Let $V = \{v_l\}_{l=1}^k$ be the vector of menu options on day $i$. Let $S$ be the polygon approximation of $\tilde{R}_{\geq i+1}$ defined by the points in $V \cup [n+i, v_{\max}]$.  We have that the revenue generated by $M$ day $i+1$ onwards is at most

 \[\sum_{j=0}^{v_{\max}} a_i(j) \tilde{R}_{\geq i+1}(\min(j,3n+i+1))   = \sum_{j=0}^{v_{\max}} a_i(j) S(\min(j,3n+i+1)) .\]
 
 Since for all $i$, we have $B_i \succeq A_i$, and $S(\min(3n+i+1,x))$ is increasing and concave, it holds by the $3^{\text{rd}}$ bullet in \autoref{thm:sosd} that 
 
 \[ \sum_{j=0}^{v_{\max}} a_i(j) S(\min(j,3n+i+1))  \leq  \sum_{j=0}^{v_{\max}} b_i(j) S(\min(j,3n+i+1)) .\]

 Consider the mechanism that mimics $M_B$ on days $1$ through $i$ and does optimally thereafter. Since it mimics $M_B$ on days $1$ through $i$, the revenue generated on these days is at least that of $M_A$ ($4^{\text{th}}$ bullet in \autoref{cor:cleaning}). Day $i+1$ onwards, this mechanism generates a revenue of $\sum_{j=0}^{v_{\max}} b_i(j) \tilde{R}_{\geq i+1}(\min(j,3n+i+1))$. To prove our result it is sufficient to prove that $\sum_{j=0}^{v_{\max}} b_i(j) \tilde{R}_{\geq i+1}(\min(j,3n+i+1))-\sum_{j=0}^{v_{\max}} b_i(j) S(\min(j,3n+i+1)) \geq \frac{1}{60000}  \geq \frac{\mathsf{OPT}}{200000n^2} $. Observe that since the functions $S$ and $\tilde{R}_{\geq i+1}$ are the same for all values at least $n+i$, we have

 %$\sum_{j=0}^{v_{\max}} b_i(j) \tilde{R}_{\geq i+1}(\min(j,3n+i+1))-\frac{\mathsf{OPT}}{100000n^2} \geq \sum_{j=0}^{v_{\max}} b_i(j) \tilde{R}_{\geq i+1}(\min(j,3n+i+1)) -\frac{1}{30000}  \geq \sum_{j=0}^{v_{\max}} b_i(j) S(\min(j,3n+i+1))$
  
  \[  \sum_{j=0}^{v_{\max}} b_i(j) \tilde{R}_{\geq i+1}(\min(j,3n+i+1))  - \sum_{j=0}^{v_{\max}} b_i(j) S(\min(j,3n+i+1)) =  \sum_{j=0}^{n+i} b_i(j) \left( \tilde{R}_{\geq i+1}(j) -  S(j) \right).   \]
  
  We know that $M_B$ has revenue at least that of $M_A$ and is clean. Thus, by \autoref{thm:cleanauctions} for $M_B$, we get that $b_i$ has a mass of at least $1/300n$ on at least $n/4$ points in the range $[n, n+i]$. The function $S_i$ in this range is defined by $V \cap [0, n+i]$ which has at most $n/8$ points. Thus, there are at least $n/8$ points in the support of $b_i$ that are not in $V \cap [0, n+i]$. For any such point $x$,
  \begin{equation}\label{eq:gap}
  \begin{aligned}
  S(x) &\leq \frac{1}{2}\left(R_{\geq i+1}(x-1)  + R_{\geq i+1}(x+1)\right) \\ 
  &= \frac{1}{2}\left((n-i)S_{x-n-1}  + (n-i)S_{x-n+1}\right)  &\text{(\autoref{eq:tildergeqi})}\\
  &= \frac{n-i}{2}\left( 2S_{x-n} + \frac{1}{\lambda^{x-n}} - \frac{1}{\lambda^{x-n-1}}\right)  \\
  &= (n-i)S_{x-n} - \frac{n-i}{2} \frac{\lambda - 1}{\lambda^{x-n}}   \\
  &= R_{\geq i+1}(x) - \frac{n-1}{8n} \frac{1}{\lambda^{x-n}}   &\text{(\autoref{eq:tildergeqi})}\\
  &\leq R_{\geq i+1}(x) - \frac{3(n-i)}{32n} < \frac{3}{64}.  &\text{(\autoref{claim:lambdapowers})}\\
  \end{aligned}
  \end{equation}
  
  Thus, the expression $\sum_{j=0}^{n+i} b_i(j) \left( \tilde{R}_{\geq i+1}(j) -  S(j) \right)$ is at least the same sum over the $n/8$ points where \autoref{eq:gap} holds.  This sum is thus, at least $\frac{n}{8}\cdot \frac{3}{64}\cdot \frac{1}{300n} > \frac{1}{60000}$.

 \end{proof}

\end{appendices}

\bibliographystyle{alpha}
\bibliography{MasterBib}

\end{document}